\documentclass{aa}
\usepackage{txfonts}
\usepackage{graphicx}
\usepackage{float}
\usepackage{natbib}
\usepackage[english]{babel}

\title{\textit{Spitzer} deep and wide Legacy mid- and far-infrared number counts and lower limits of cosmic infrared background}

\author{M. B\'ethermin\inst{1} \and H. Dole\inst{1} \and A. Beelen\inst{1} \and H. Aussel\inst{2}}

\institute{
\inst{1} Institut d'Astrophysique Spatiale (IAS), bat121, F-91405 Orsay, France; Universit\'e Paris-Sud 11 and CNRS (UMR8617)\\
\inst{2}Laboratoire AIM, CEA/DSM-CNRS-Universit\'e Paris Diderot, IRFU/Service d'Astrophysique, B‰t. 709, CEA-Saclay, 91191 Gif-sur-Yvette Cedex, France
}

\date{Received 11 September 2009 / Accepted 05 January 2009}

\abstract{}{
We aim to place stronger lower limits on the cosmic infrared background (CIB) brightness at 24~$\mu$m, 70~$\mu$m and 160~$\mu$m and measure the extragalactic number counts at these wavelengths in a homogeneous way from various surveys.
}{
Using \textit{Spitzer} legacy data over 53.6 $deg^2$ of various depths, we build catalogs with the same extraction method at each wavelength. Completeness and photometric accuracy are estimated with Monte-Carlo simulations. Number count uncertainties are estimated with a counts-in-cells moment method to take galaxy clustering into account. Furthermore, we use a stacking analysis to estimate number counts of sources not detected at 70~$\mu$m and 160 $\mu$m. This method is validated by simulations. The integration of the number counts gives new CIB lower limits.
}{
Number counts reach 35~$\mu$Jy, 3.5~mJy and 40~mJy at 24~$\mu$m, 70~$\mu$m, and 160~$\mu$m, respectively. We reach deeper flux densities of 0.38 mJy at 70, and 3.1 at 160 $\mu$m with a stacking analysis. We confirm the number count turnover at 24~$\mu$m and 70~$\mu$m, and observe it for the first time at 160~$\mu$m at about 20~mJy, together with a power-law behavior below 10~mJy. These mid- and far-infrared counts: 1) are homogeneously built by combining fields of different depths and sizes, providing a legacy over about three orders of magnitude in flux density; 2) are the deepest to date at 70~$\mu$m and 160~$\mu$m;  3) agree with previously published results in the common measured flux density range; 4) globally agree with the Lagache et al. (2004) model, except at 160 $\mu$m, where the model slightly overestimates the counts around 20 and 200 mJy.
}{
These counts are integrated to estimate new CIB firm lower limits of $2.29_{-0.09}^{+0.09}$~nW.m$^{-2}$.sr$^{-1}$, $5.4_{-0.4}^{+0.4}$~nW.m$^{-2}$.sr$^{-1}$, and $8.9_{-1.1}^{+1.1}$~nW.m$^{-2}$.sr$^{-1}$ at 24~$\mu$m, 70~$\mu$m, and 160~$\mu$m, respectively, and extrapolated to give new estimates of the CIB due to galaxies of $2.86_{-0.16}^{+0.19}$~nW.m$^{-2}$.sr$^{-1}$, $6.6_{-0.6}^{+0.7}$~nW.m$^{-2}$.sr$^{-1}$, and $14.6_{-2.9}^{+7.1}$~nW.m$^{-2}$.sr$^{-1}$, respectively. Products (point spread function, counts, CIB contributions, software) are publicly available for download at http://www.ias.u-psud.fr/irgalaxies/.
}

\titlerunning{\textit{Spitzer} deep wide legacy MIR/FIR counts and CIB lower limits}
\authorrunning{B\'ethermin et al.}

\begin{document}

\maketitle

\begin{table*}
\centering
\begin{tabular}{|l|rrr|rrr|rrr|}
\hline
Field name & \multicolumn{3}{|c|}{Surface area} & \multicolumn{3}{|c|}{80\% completeness flux} & \multicolumn{3}{|c|}{Scaling factor} \\
\cline{2-10}
 & 24 $\mu m$  & 70 $\mu m$ & 160 $\mu m$ & 24 $\mu m$ & 70 $\mu m$ & 160 $\mu m$ & 24 $\mu m$ & 70 $\mu m$ & 160 $\mu m$\\
\cline{2-10}
 & \multicolumn{3}{|c|}{deg$^2$} & $\mu Jy$ & \multicolumn{2}{|c|}{$mJy$} & \multicolumn{3}{|c|}{~}\\
\hline
FIDEL eCDFS &  0.23 &  0.19 & - &  60. &  4.6 & - & 1.0157 & 1 & - \\
FIDEL EGS &  0.41 & - &  0.38 &  76. & - &  45. & 1.0157 & - & 0.93 \\
COSMOS &  2.73 &  2.41 &  2.58 &  96. &  7.9 &  46. & 1 & 0.92 & 0.96 \\
SWIRE LH & 10.04 & 11.88 & 11.10 & 282. & 25.4 &  92. & 1.0509 & 1.10 & 0.93 \\
SWIRE EN1 &  9.98 &  9.98 &  9.30 & 261. & 24.7 &  94. & 1.0509 & 1.10 & 0.93 \\
SWIRE EN2 &  5.36 &  5.34 &  4.98 & 267. & 26.0 &  90. & 1.0509 & 1.10 & 0.98 \\
SWIRE ES1 &  7.45 &  7.43 &  6.71 & 411. & 36.4 & 130. & 1.0509 & 1.10 & 0.98 \\
SWIRE CDFS &  8.42 &  8.28 &  7.87 & 281. & 24.7 &  88. & 1.0509 & 1.10 & 0.98 \\
SWIRE XMM &  8.93 & - & - & 351. & - & - & 1.0509 & - & - \\
\hline
Total & 53.55 & 45.51  & 42.91 & \multicolumn{6}{|c}{~}\\
\cline{1-4}
\end{tabular}

\caption{\label{tab:fieldlist} Size, 80\% completeness flux density and calibration scaling factor (see Sect. \ref{subsection:data}) of the used fields. Some fields are not used at all wavelengths.}
\end{table*}

\section{Introduction}

The extragalactic background light (EBL) is the relic emission of all processes of structure formation in the Universe. About half of this emission, called the Cosmic Infrared Background (CIB) is emitted in the 8-1000~$\mu$m range, and peaks around 150~$\mu$m. It is essentially due to the star formation \citep{Puget1996,Fixsen1998,Hauser1998, Lagache1999, Gispert2000, Hauser2001,Kashlinsky2005, Lagache2005}.\\

The CIB spectral energy distribution (SED) is an important constraint for the infrared galaxies evolution models (e.g. \citet{Lagache2004,Franceschini2009,Le_Borgne2009,Pearson2009,Rowan2009,Valiante2009}). It gives the budget of infrared emission since the first star. The distribution of the flux of sources responsible for this background is also a critical constraint. We propose to measure the level of the CIB and the flux distribution of the sources at 3 wavelengths (24~$\mu$m, 70~$\mu$m and 160~$\mu$m).\\

In the 1980's, the infrared astronomical satellite (IRAS) and COBE/DIRBE performed the first mid-infrared (MIR) and far-infrared (FIR) full-sky surveys. Nevertheless, the detected sources were responsible for a very small part of the CIB. Between 1995 and 1998, the ISO (infrared space observatory) performed deeper observations of infrared galaxies. \citet{Elbaz1999} resolved into the source more than half of the CIB at 15~$\mu$m. At larger wavelengths, the sensitivity and angular resolution was not sufficient to resolve the CIB \citep{Dole2001}.\\

The \textit{Spitzer} space telescope \citep{Werner2004}, launched in 2003, has performed deep infrared observations on wide fields. The multiband imaging photometers for \textit{Spitzer} (MIPS) \citep{Rieke2004} mapped  the sky at 24~$\mu$m, 70~$\mu$m and 160~$\mu$m. About 60\% of the CIB was resolved at 24 $\mu$m \citep{Papovitch2004} and at 70 $\mu$m \citep{Frayer2006b}. Because of confusion \citep{Dole2003}, only about 7\% were resolved at 160 $\mu$m \citep{Dole2004}. \citet{Dole2006} managed to resolve most of the 70~$\mu$m and 160~$\mu$m by stacking 24 $\mu$m sources.\\

The cold mission of Spitzer is over, and lots of data are now public. We present extragalactic number counts built homogeneously by combining deep and wide fields. The large sky surface used significantly reduces  uncertainties on number counts. In order to obtain very deep FIR number counts, we used a stacking analysis and estimate the level of the CIB in the three MIPS bands with them.\\

\section{Data, source extraction and photometry}

\label{section:se}

\subsection{Data}

\label{subsection:data}

We took the public \textit{Spitzer} mosaics\footnote{from the Spitzer Science Center website: http://data.spitzer.caltech.edu/popular/} from different observation programs: the GOODS/FIDEL (PI: M. Dickinson), COSMOS (PI: D. Sanders) and SWIRE (PI: C. Lonsdale). We used only the central part of each field, which was defined by a cut of 50\% of the median coverage for SWIRE fields and 80 \% for the other. The total area covers 53.6~deg$^2$, 45.5~deg$^2$, 42.9~deg$^2$ at 24~$\mu$m, 70~$\mu$m and 160~$\mu$m respectively. The surface of the deep fields (FIDEL, COSMOS) is about 3.5~deg$^2$. Some fields were not used at all wavelengths for different reasons: There is no public release of FIDEL CDFS data at 160 $\mu$m; the pixels of the EGS 70 $\mu$m are not square; XMM is not observed at 70 and 160 $\mu$m. Table \ref{tab:fieldlist} summarises the field names, sizes and completenesses.\\

In 2006, new calibration factors were adopted for MIPS \citep{Engelbracht2007,Gordon2007,Stansberry2007}. The conversion factor from instrumental unit to MJy/sr is 0.0454 (resp. 702 and 41.7) at 24~$\mu$m (resp. 70~$\mu$m and 160~$\mu$m). The COSMOS GO3 and SWIRE (released 22 Dec. 2006) mosaics were generated with the new calibration. The FIDEL mosaics were obtained with other factors at 24~$\mu$m and 160~$\mu$m (resp. 0.0447 and 44.7). The 70~$\mu$m and 160~$\mu$m COSMOS mosaics were color corrected (see Sect. \ref{subsection:colcor}). Consequently we applied a scaling factor (see Table \ref{tab:fieldlist}) before the source extraction to each mosaic to work on a homogeneous sample of maps (new calibration and no color correction).

\subsection{Source extraction and photometry}

The goal is to build homogeneous number counts with well-controlled systematics and high statistics. However, the fields present various sizes and depths. We thus employed a single extraction method at a given wavelength, allowing the heterogeneous datasets to combine in a coherent way.\\

\subsubsection{Mid-IR/far-IR differences}

The MIR (24 $\mu$m) and FIR (70~$\mu$m and 160~$\mu$m) maps have different properties: in the MIR, we observe lots of faint blended sources; in the FIR, due to confusion \citep{Dole2004}, all these faint blended sources are only seen as background fluctuations. Consequently, we used different extraction and photometry methods for each wavelength. In the MIR, the priority is the deblending: accordingly we took the SExtractor \citep{Bertin1996} and PSF fitting. In the FIR, we used efficient methods with strong background fluctuations: wavelet filtering, threshold detection and aperture photometry.\\

\subsubsection{Point spread function (PSF)}

The 24 $\mu$m empirical PSF of each field is generated with the IRAF (image reduction and analysis facility\footnote{http://iraf.noao.edu/}) DAOPHOT package \citep{Stetson1987} on the 30 brightest sources of each map. It is normalized in a 12 arcsec radius aperture. Aperture correction (1.19) is computed with the S Tiny Tim\footnote{http://ssc.spitzer.caltech.edu/archanaly/contributed/stinytim/}  \citep{Krist2006} theoretical PSF for a constant $\nu S_{\nu}$ spectrum. The difference of correction between a $S_{\nu} = \nu^{-2}$ and a $\nu^2$ spectrum is less than 2\%. So, the hypothesis on the input spectrum is not critical for the PSF normalization.\\

At 70~$\mu$m and 160~$\mu$m, we built a single empirical PSF from the SWIRE fields. We used the Starfinder PSF extraction routine \citep{Diolaiti2000}, which median-stacks the brightest non-saturated sources (100~mJy~$<S_{70}<10$~Jy and 300~mJy$<S_{160}<1$~Jy). Previously, fainter neighboring sources were subtracted with a first estimation of the PSF. At 70~$\mu$m (resp. 160~$\mu$m), the normalization is done in a 35~arcsec (resp. 80~arcsec) aperture, with a sky annulus between 75~arcsec and 125~arcsec (resp. 150~arcsec and 250~arcsec); the aperture correction was 1.21 (resp. 1.20). The theoretical signal in the sky annulus and the aperture correction were computed with the S Tiny Tim Spitzer PSF for a constant $\nu S_{\nu}$ spectrum. These parameters do not vary more than 5 \% with the spectrum of sources. Pixels that were affected by the temporal median filtering artifact, which was sometimes present around bright sources, were masked prior to these operations.\\

\subsubsection{Source extraction and photometry}

At 24 $\mu$m, we detected sources with SExtractor. We chose a Gaussian filter (gauss\_5.0\_9x9.conv) and a background filter of the size of $64\times64$ pixels. The detection and analysis thresholds were tuned for each field. We performed PSF fitting photometry with the  DAOPHOT allstar routine. This routine is very efficient for blended sources flux measurement.\\

At 70~$\mu$m and~160 $\mu$m, we applied the a-trou wavelet filtering \citep{Starck1999} on the maps to remove the large scale fluctuations (10 pixels) on which we performed the source detection with a threshold algorithm \citep{Dole2001, Dole2004}. The threshold was tuned for each field. Photometry was done by aperture photometry on a non filtered map at the positions found on the wavelet filtered map. At 70 $\mu$m, we used 10 arcsec aperture radius and a 18 arcsec to 39 arcsec sky annulus. At 160 $\mu$m, we used an aperture of 20 arcsec and a 40 arcsec to 75 arcsec annulus. Aperture corrections were computed with the normalized empirical PSF: 3.22 at 70~$\mu$m and 3.60 at 160~$\mu$m. In order to estimate the uncertainty on this correction, aperture corrections were computed using five PSF built on five different SWIRE fields. The uncertainty is 1.5\% at 70~$\mu$m and 4.5\% at 160~$\mu$m.\\

\subsection{Color correction}

\label{subsection:colcor}

The MIPS calibration factors were calculated for a 10000 K blackbody (MIPS Data Handbook 2007\footnote{http://ssc.spitzer.caltech.edu/mips/dh/}). However, the galaxies SED (spectral energy distribution) are different and the MIPS photometric bands are large ($\lambda / \Delta \lambda \approx 3$). Thus, color corrections were needed. We used (like \citet{Shupe2008} and \citet{Frayer2009}) a constant $\nu S_{\nu}$ spectrum at 24~$\mu$m, 70~$\mu$m and 160~$\mu$m. Consequently, all fluxes were divided by 0.961, 0.918 and 0.959 at 24~$\mu$m, 70~$\mu$m and 160~$\mu$m due to this color correction. Another possible convention is $\nu S_{\nu} \propto \nu^{-1}$. This convention is more relevant for the local sources at 160~$\mu$m, whose spectrum decreases quickly with wavelength. Nevertheless, the redshifted sources studied by stacking are seen at their peak of the cold dust emission, and their SED agrees better with the constant $\nu S_{\nu}$ convention. The difference of color correction between these two conventions is less than 2 \%, and this choice is thus not critical. We consequently chose the constant $\nu S_{\nu}$ convention to more easily compare our results with \citet{Shupe2008} and \citet{Frayer2009}.\\

\section{Catalog properties}

\subsection{Spurious sources}

\label{subsection:spurious}

\begin{figure}
\centering
\includegraphics{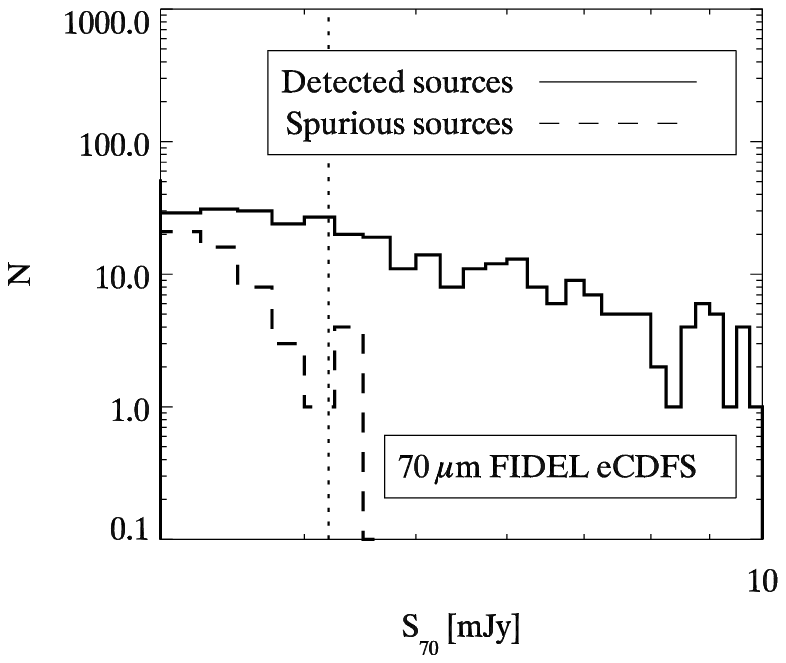}[h]
\caption{\label{fig:spurious}Flux distribution of sources extracted from normal \textit{(solid line)} and flipped \textit{(dash line)} maps, at 70 $\mu$m in FIDEL eCDFS. The vertical dashed line represents the 80\% completeness flux density.}
\end{figure}

Our statistical analysis may suffer from spurious sources. We have to estimate how many false detections are present in a map and what their flux distribution is. To do so, we built a catalog with the flipped map. To build this flipped map, we multiplied the values of the pixels of the original map by a factor of -1. Detection and photometry parameters were exactly the same as for normal catalogs. At 24 $\mu$m, there are few spurious sources ($<10\%$) in bins brighter than the 80\% completeness limit flux density. At 70~$\mu$m and 160~$\mu$m, fluctuations of the background due to unresolved faint sources are responsible for spurious detections. Nevertheless, the ratio between detected source numbers and spurious source numbers stayed reasonable (below 0.2) down to the 80\% completeness limit (see the example of FIDEL CDFS at 70 $\mu$m in Fig. \ref{fig:spurious}).\\

\subsection{Completeness}

\begin{figure*}
\centering
\includegraphics{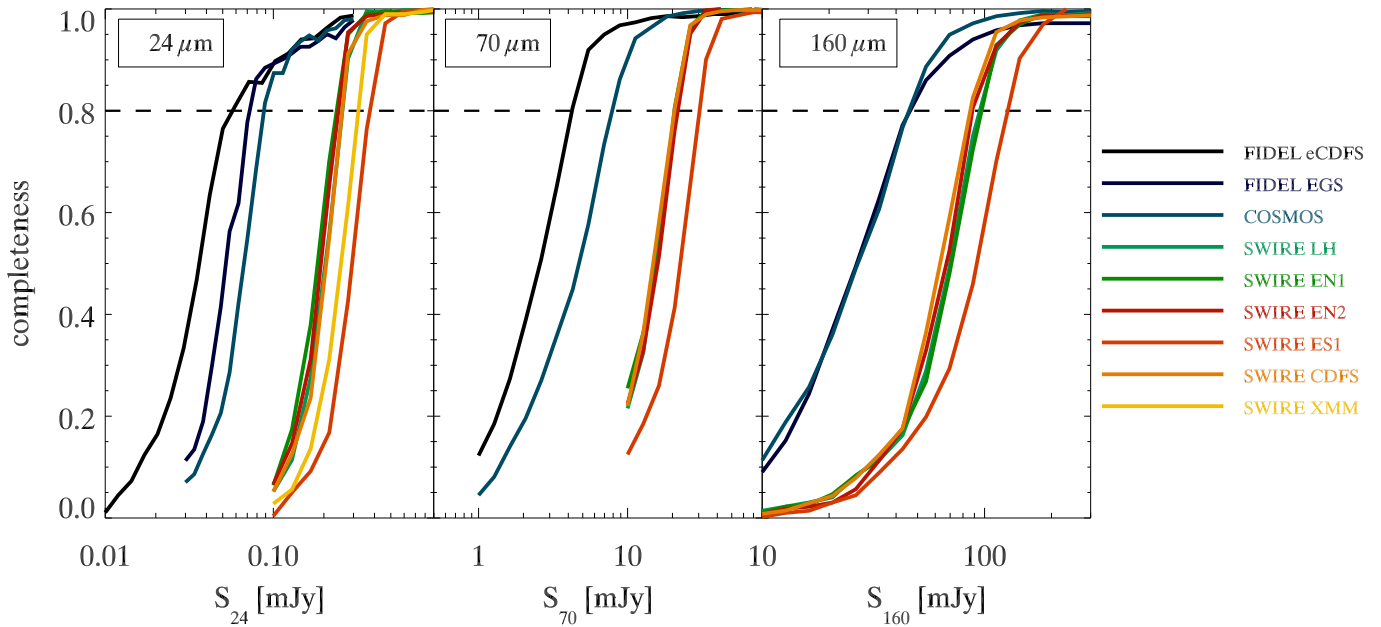}
\caption{\label{fig:comp}Completeness at 24~$\mu$m \textit{(left)}, 70~$\mu$m \textit{(center)}, and 160~$\mu$m \textit{(right)} as a function of the source flux for all fields. The dashed line represents 80\% completeness.}
\end{figure*}

The completeness is the probability to extract a source of a given flux. To estimate it, we added artificial sources (based on empirical PSF) on the initial map and looked for a detection in a 2 arcsec radius at 24~$\mu$m around the initial position (8 arcsec at 70~$\mu$m and 16 arcsec at 160~$\mu$m). This operation was done for different fluxes with a Monte-Carlo simulation. We chose the number of artificial sources in each realization in a way that they have less than 1\% probability to fall at a distance shorter than 2 PSF FWHM (full width at half maximum). The completeness is plotted in Fig. \ref{fig:comp}, and the 80\% completeness level is reported in Table \ref{tab:fieldlist}.\\

\subsection{Photometric accuracy}

\label{subsection:photo_acc}

\begin{figure}
\centering
\includegraphics{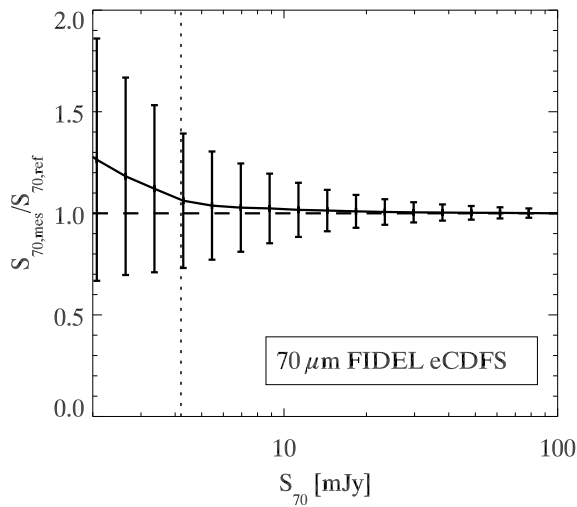}[h]
\caption{\label{fig:photcal}Ratio between measured flux and input flux computed from Monte Carlo simulations at 70~$\mu$m in FIDEL eCDFS. Error bars represent 1 $\sigma$ dispersion. The vertical dashed line represents the 80\% completeness flux density.}
\end{figure}

The photometric accuracy was checked with the same Monte-Carlo simulation. For different input fluxes, we built histograms of measured fluxes and computed the median and scatter of these distributions. At lower fluxes, fluxes are overestimated and errors are larger. These informations were used to estimate the Eddington bias (see next section). The photometric accuracy at 70 $\mu$m in FIDEL CDFS is plotted as an example in Fig. \ref{fig:photcal}.\\

We also compared our catalogs with published catalogs. At 24 $\mu$m, we compared it with the GOODS CDFS catalog of \citet{Chary2004}, and the COSMOS catalog of \citet{Le_Floch2009}. Their fluxes were multiplied by a corrective factor to be compatible with the $\nu S_{\nu}$ = constant convention. Sources were considered to be the same if they are separated by less than 2 arcsec. We computed the standard deviation of the distribution of the ratio between our and their catalogs. In a 80-120 $\mu Jy$ bin in the CDFS, we found a dispersion of 19\%. In a 150-250 $\mu Jy$ bin in COSMOS, we found a scatter of 13\%. The offset is +3\% with COSMOS catalog and -1\% with GOODS catalog. At 70 and 160 $\mu$m, we compared our catalogs with the COSMOS and SWIRE team ones. In all cases, the scatter is less than 15\%, and the offset is less than 3\%. At all wavelengths and for all fields, the offset is less than the calibration uncertainty.\\

\subsection{Eddington bias}

\label{subsection:edd_bias}

\begin{figure}[h]
\centering
\includegraphics{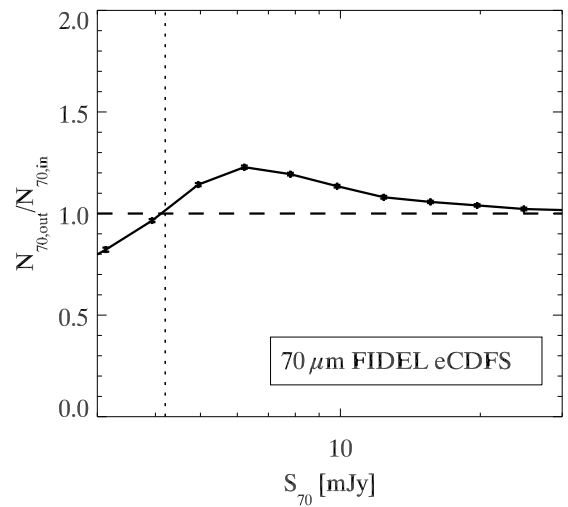}
\caption{\label{fig:eddbias}Eddington bias: ratio between the number of detected sources and the number of input sources at 70~$\mu$m in FIDEL eCDFS. The vertical dashed line represents the 80\% completeness flux density.}
\end{figure}

When sources become fainter, photometric errors increase. In addition, fainter sources are more numerous  than brighter ones (in general $dN/dS \sim S^{-r}$). Consequently, the number of sources in faint bins are overestimated. This is the classical Eddington bias \citep{Eddington1913,Eddington1940}. The example of FIDEL CDFS at 70 $\mu$m is plotted in Fig. \ref{fig:eddbias}.\\

To correct for this effect at 70~$\mu$m and 160~$\mu$m, we estimated a correction factor for each flux bin. We generated an input flux catalog with a power-law distribution (r = 1.6 at 70~$\mu$m, r = 3 at 160~$\mu$m). We took into account completeness and photometric errors (coming from Monte-Carlo simulations) to generate a mock catalog. We then computed the ratio between the number of mock sources found in a bin and the number of input sources. This task was done for all fields. This correction is more important for large r (at 160 $\mu$m). At 24 $\mu$m, thanks to the PSF fitting, the photometric error is more reduced and symmetrical. Less faint sources are thus placed in brighter flux bins. Because of this property and the low r (about 1.45), this correction can be ignored for 24 $\mu$m counts.\\

\section{Number counts}
\label{section:counts}

\subsection{Removing stars from the catalogs}

To compute extragalactic number counts at 24~$\mu$m, we removed the stars with the $K-[24] < 2$ color criterion and identification procedure following \citet{Shupe2008}. The K band magnitudes were taken from the 2MASS catalog \citep{Skrutskie2006}. We ignored the star contribution at 70~$\mu$m and 160~$\mu$m, which is negligible ($<$1\% in all used flux density bins) according to the DIRBE Faint Count model \citep{Arendt1998}.\\

\subsection{24 $\mu$m number counts}

\begin{figure*}
\centering
\includegraphics{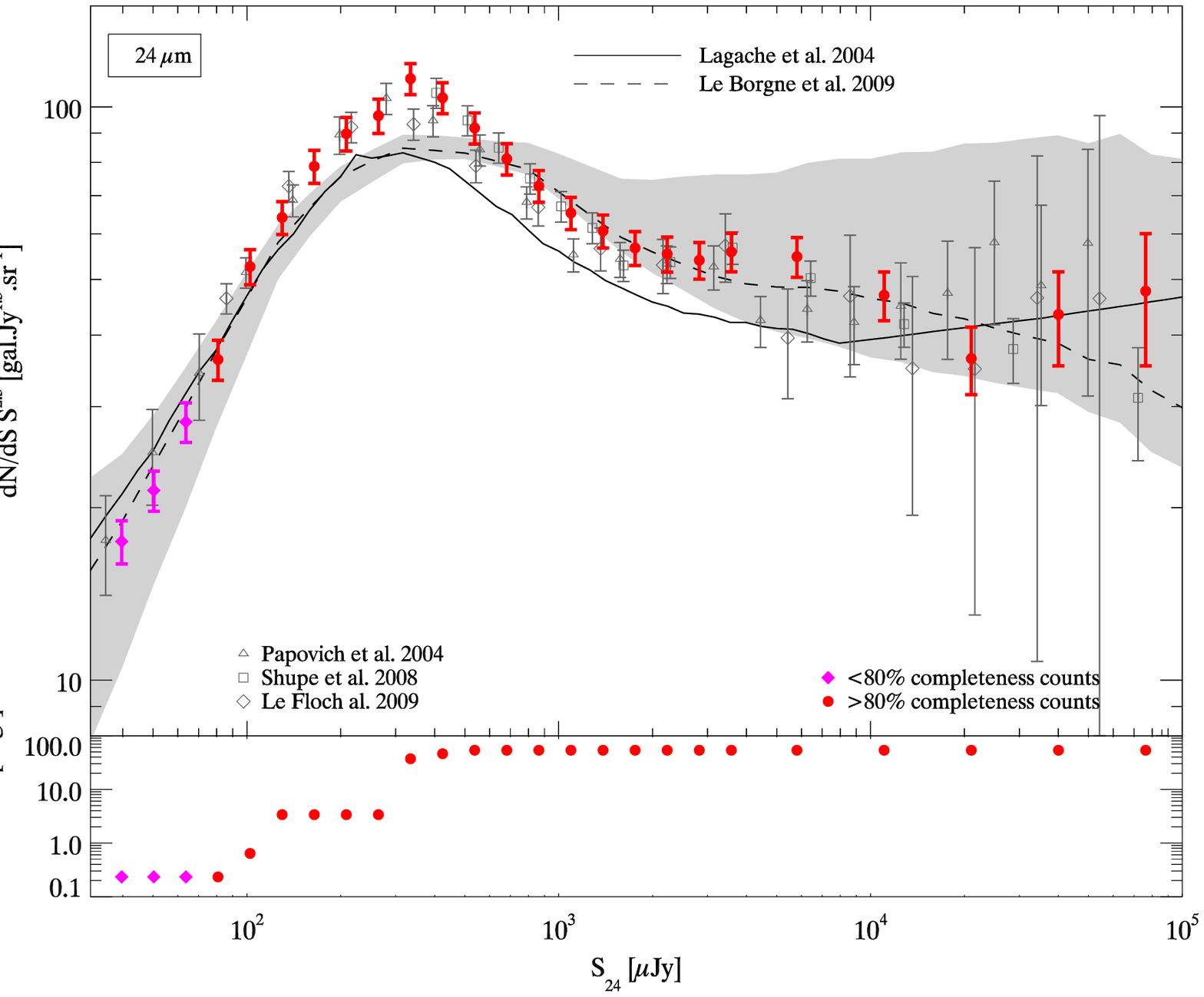}[h]
\caption{Differential number counts at 24~$\mu$m. \textit{Filled circle}: points obtained with $\ge$80\% completeness; \textit{filled diamond}: points obtained with a 50\% to 80\% completeness; \textit{open triangle}: \citet{Papovitch2004} GTO number counts obtained with PSF fitting photometry; \textit{open square}: \citet{Shupe2008} SWIRE number counts obtained with aperture photometry; \textit{open diamond}: \citet{Le_Floch2009} COSMOS number counts obtained with PSF fitting photometry; \textit{continuous line}: \citet{Lagache2004} model; \textit{dashed line and grey region}: \citet{Le_Borgne2009} model and 90\% confidence region. Error bars take into accounts clustering (see section \ref{subsection:sectclus}) and calibration uncertainties \citep{Engelbracht2007}.}
\label{fig:24nc}
\end{figure*}

\begin{table*}
\centering
\begin{tabular}{|rrr|r|rrr|r|}
\hline
$<S>$ & $S_{min}$ & $S_{max}$ & $dN/dS.S^{2.5}$ & $\sigma_{poisson}$ & $\sigma_{clustering}$ & $\sigma_{clus.+calib.}$ & $\Omega_{used}$\\
\hline
\multicolumn{3}{|c|}{(in mJy)} & \multicolumn{4}{|c|}{(in gal.Jy$^{1.5}$.sr$^{-1}$)} & deg$^2$\\
\hline
   0.040 &    0.035 &    0.044 &   17.5 &    1.0 &    1.1 &    1.3 &   0.2 \\
   0.050 &    0.044 &    0.056 &   21.4 &    1.0 &    1.1 &    1.4 &   0.2 \\
   0.064 &    0.056 &    0.071 &   28.2 &    1.2 &    1.5 &    1.8 &   0.2 \\
   0.081 &    0.071 &    0.090 &   36.2 &    1.5 &    1.9 &    2.4 &   0.2 \\
   0.102 &    0.090 &    0.114 &   52.6 &    1.3 &    1.9 &    2.9 &   0.6 \\
   0.130 &    0.114 &    0.145 &   64.1 &    1.0 &    1.7 &    3.1 &   3.4 \\
   0.164 &    0.145 &    0.184 &   78.7 &    1.1 &    2.2 &    3.8 &   3.4 \\
   0.208 &    0.184 &    0.233 &   89.8 &    1.3 &    2.8 &    4.5 &   3.4 \\
   0.264 &    0.233 &    0.295 &   96.5 &    1.5 &    3.3 &    5.1 &   3.4 \\
   0.335 &    0.295 &    0.374 &  112.0 &    0.8 &    1.8 &    4.8 &  37.2 \\
   0.424 &    0.374 &    0.474 &  103.7 &    0.6 &    1.7 &    4.5 &  46.1 \\
   0.538 &    0.474 &    0.601 &   91.9 &    0.6 &    1.5 &    4.0 &  53.6 \\
   0.681 &    0.601 &    0.762 &   81.2 &    0.6 &    1.5 &    3.6 &  53.6 \\
   0.863 &    0.762 &    0.965 &   72.8 &    0.7 &    1.6 &    3.3 &  53.6 \\
   1.094 &    0.965 &    1.223 &   65.3 &    0.8 &    1.6 &    3.1 &  53.6 \\
   1.387 &    1.223 &    1.550 &   60.8 &    0.9 &    1.7 &    3.0 &  53.6 \\
   1.758 &    1.550 &    1.965 &   56.7 &    1.0 &    1.8 &    2.9 &  53.6 \\
   2.228 &    1.965 &    2.490 &   55.4 &    1.2 &    2.1 &    3.0 &  53.6 \\
   2.823 &    2.490 &    3.156 &   54.0 &    1.5 &    2.3 &    3.2 &  53.6 \\
   3.578 &    3.156 &    4.000 &   55.9 &    1.8 &    2.7 &    3.5 &  53.6 \\
   5.807 &    4.000 &    7.615 &   54.8 &    1.5 &    2.9 &    3.6 &  53.6 \\
  11.055 &    7.615 &   14.496 &   46.9 &    2.3 &    3.6 &    4.1 &  53.6 \\
  21.045 &   14.496 &   27.595 &   36.4 &    3.3 &    4.4 &    4.6 &  53.6 \\
  40.063 &   27.595 &   52.531 &   43.4 &    5.9 &    7.7 &    7.9 &  53.6 \\
  76.265 &   52.531 &  100.000 &   47.7 &    9.9 &   12.0 &   12.2 &  53.6 \\
\hline
\end{tabular}
\caption{\label{tab:24chart} Differential number counts at 24 $\mu$m. $\sigma_{clustering}$ is the uncertainty taking into account clustering (see Sect. \ref{subsection:sectclus}). $\sigma_{clus.+calib.}$ takes into account both clustering and calibration \citep{Engelbracht2007}.}

\begin{tabular}{|rrr|r|rrr|r|}
\hline
$<S>$ & $S_{min}$ & $S_{max}$ & $dN/dS.S^{2.5}$ & $\sigma_{poisson}$ & $\sigma_{clustering}$ & $\sigma_{clus.+calib.}$ & $\Omega_{used}$\\
\hline
\multicolumn{3}{|c|}{(in mJy)} & \multicolumn{4}{|c|}{(in gal.Jy$^{1.5}$.sr$^{-1}$)} & deg$^2$\\
\hline
   4.197 &    3.500 &    4.894 &  2073. &   264. &   309. &   342. &   0.2 \\
   5.868 &    4.894 &    6.843 &  2015. &   249. &   298. &   330. &   0.2 \\
   8.206 &    6.843 &    9.569 &  1690. &   289. &   332. &   353. &   0.2 \\
  11.474 &    9.569 &   13.380 &  2105. &   123. &   202. &   250. &   2.6 \\
  16.044 &   13.380 &   18.708 &  2351. &   148. &   228. &   281. &   2.6 \\
  22.434 &   18.708 &   26.159 &  1706. &   153. &   208. &   240. &   2.6 \\
  31.369 &   26.159 &   36.578 &  2557. &    69. &   124. &   218. &  38.1 \\
  43.862 &   36.578 &   51.146 &  2446. &    73. &   123. &   211. &  45.5 \\
  61.331 &   51.146 &   71.517 &  2359. &    90. &   141. &   217. &  45.5 \\
  85.758 &   71.517 &  100.000 &  2257. &   112. &   164. &   228. &  45.5 \\
 157.720 &  100.000 &  215.440 &  2354. &   121. &   198. &   257. &  45.5 \\
 339.800 &  215.440 &  464.160 &  2048. &   200. &   276. &   311. &  45.5 \\
 732.080 &  464.160 & 1000.000 &  2349. &   381. &   500. &   526. &  45.5 \\
\hline
\end{tabular}
\caption{\label{tab:70chart} Differential number counts at 70 $\mu$m. $\sigma_{clustering}$ is the uncertainty taking into account clustering (see Sect. \ref{subsection:sectclus}). $\sigma_{clus.+calib.}$ takes into account both clustering and calibration \citep{Gordon2007}.}

\begin{tabular}{|rrr|r|rrr|r|}
\hline
$<S>$ & $S_{min}$ & $S_{max}$ & $dN/dS.S^{2.5}$ & $\sigma_{poisson}$ & $\sigma_{clustering}$ & $\sigma_{clus.+calib.}$ & $\Omega_{used}$\\
\hline
\multicolumn{3}{|c|}{(in mJy)} & \multicolumn{4}{|c|}{(in gal.Jy$^{1.5}$.sr$^{-1}$)} & deg$^2$\\
\hline
  45.747 &   40.000 &   51.493 & 16855. &  1312. &  2879. &  3519. &   3.0 \\
  58.891 &   51.493 &   66.289 & 14926. &  1243. &  2704. &  3243. &   3.0 \\
  75.813 &   66.289 &   85.336 & 13498. &  1319. &  2648. &  3104. &   3.0 \\
  97.596 &   85.336 &  109.860 & 12000. &  1407. &  2442. &  2835. &   3.0 \\
 125.640 &  109.860 &  141.420 & 10687. &   457. &   991. &  1621. &  36.2 \\
 161.740 &  141.420 &  182.060 &  7769. &   425. &   773. &  1211. &  42.9 \\
 208.210 &  182.060 &  234.370 &  7197. &   472. &   810. &  1184. &  42.9 \\
 268.040 &  234.370 &  301.710 &  5406. &   487. &   734. &   979. &  42.9 \\
 345.050 &  301.710 &  388.400 &  5397. &   585. &   843. &  1063. &  42.9 \\
 444.200 &  388.400 &  500.000 &  4759. &   662. &   891. &  1059. &  42.9 \\
 750.000 &  500.000 & 1000.000 &  6258. &   685. &  1158. &  1380. &  42.9 \\
1500.000 & 1000.000 & 2000.000 &  4632. &   989. &  1379. &  1487. &  42.9 \\
\hline
\end{tabular}
\caption{\label{tab:160chart} Differential number counts at 160 $\mu$m. $\sigma_{clustering}$ is the uncertainty taking into account clustering (see Sect. \ref{subsection:sectclus}). $\sigma_{clus.+calib.}$ takes into account both clustering and calibration \citep{Stansberry2007}.}

\end{table*}

We counted the number of extragalactic sources for each field and in each flux bin. We subtracted the number of spurious detections (performed on the flipped map). We divided by the completeness. As a next step, the counts of all fields were combined together with a mean weighted by field size. Actually, a weighting by the number of sources in each field overweighs the denser fields and biases the counts. Counts from a field were combined only if the lower end of the flux bin was larger then or equal to the 80\% completeness. We thus reached  71~$\mu$Jy (71~$\mu$Jy to 90~$\mu$Jy bin) in the counts. However, to probe fainter flux densities, we used the data from the deepest field (FIDEL eCDFS) between a 50 and 80 \% completeness, allowing us to reach 35~$\mu$Jy.\\

Our number counts are plotted in Fig. \ref{fig:24nc} and are written in Table \ref{tab:24chart}. We also plot data from \citet{Papovitch2004}, \citet{Shupe2008} and \citet{Le_Floch2009}, and model predictions from \citet{Lagache2004} and \citet{Le_Borgne2009}. The \citet{Papovitch2004} fluxes are multiplied by a factor 1.052 to take into account the update in the calibration, the color correction and the PSF. This correction of flux also implies a correction on number counts, according to:
\begin{equation}
\big ( \frac{dN}{dS_{f}} S_{f}^{2.5} \big )_{S_{f}} = \big ( c^{1.5} \frac{dN}{dS_{i}} S_{i}^{2.5} \big )_{c S_{i}},
\end{equation}
where $S_{i}$ is the initial flux, $S_{f}$ is the corrected flux and c the corrective factor ($S_{f} = c S_{i}$). A correction of the flux thus corresponds to a shift in the abscissa (factor $c$) and in the ordinate (factor $c^{1.5}$). \citet{Papovitch2004} do not subtract stars and thus overestimate counts above 10 mJy. We have a very good agreement with their work below 10~mJy. We also have a very good agreement with \citet{Shupe2008}. The \citet{Le_Floch2009} fluxes are multiplied by 1.05 to take into account a difference of the reference SED: 10~000K versus constant $\nu S_{\nu}$, and by another correction of 3\% corresponding to the offset observed in Sect. \ref{subsection:photo_acc}. There is an excellent agreement with their work.\\

The \citet{Lagache2004}\footnote{\citet{Lagache2004} model used a $\Lambda$CDM cosmology with $\Omega_{\Lambda}$=0.73, $\Omega_{M}$=0.27 and h = 0.71} and \citet{Le_Borgne2009}\footnote{\citet{Le_Borgne2009} model used a $\Lambda$CDM cosmology with $\Omega_{\Lambda}$=0.7, $\Omega_{M}$=0.3 and h = 0.7} generally agree well with the data, in particular on the faint end below 100 $\mu$Jy, and on the position of the peak around 300 $\mu$Jy. However, the \citet{Lagache2004} model slightly underestimates (about 10\%) the counts above 200 $\mu$Jy. The \cite{Le_Borgne2009} model is flatter than the data, and agrees reasonably well above 600$\mu$Jy.\\

\subsection{70 $\mu$m number counts}

\begin{figure*}
\centering
\includegraphics{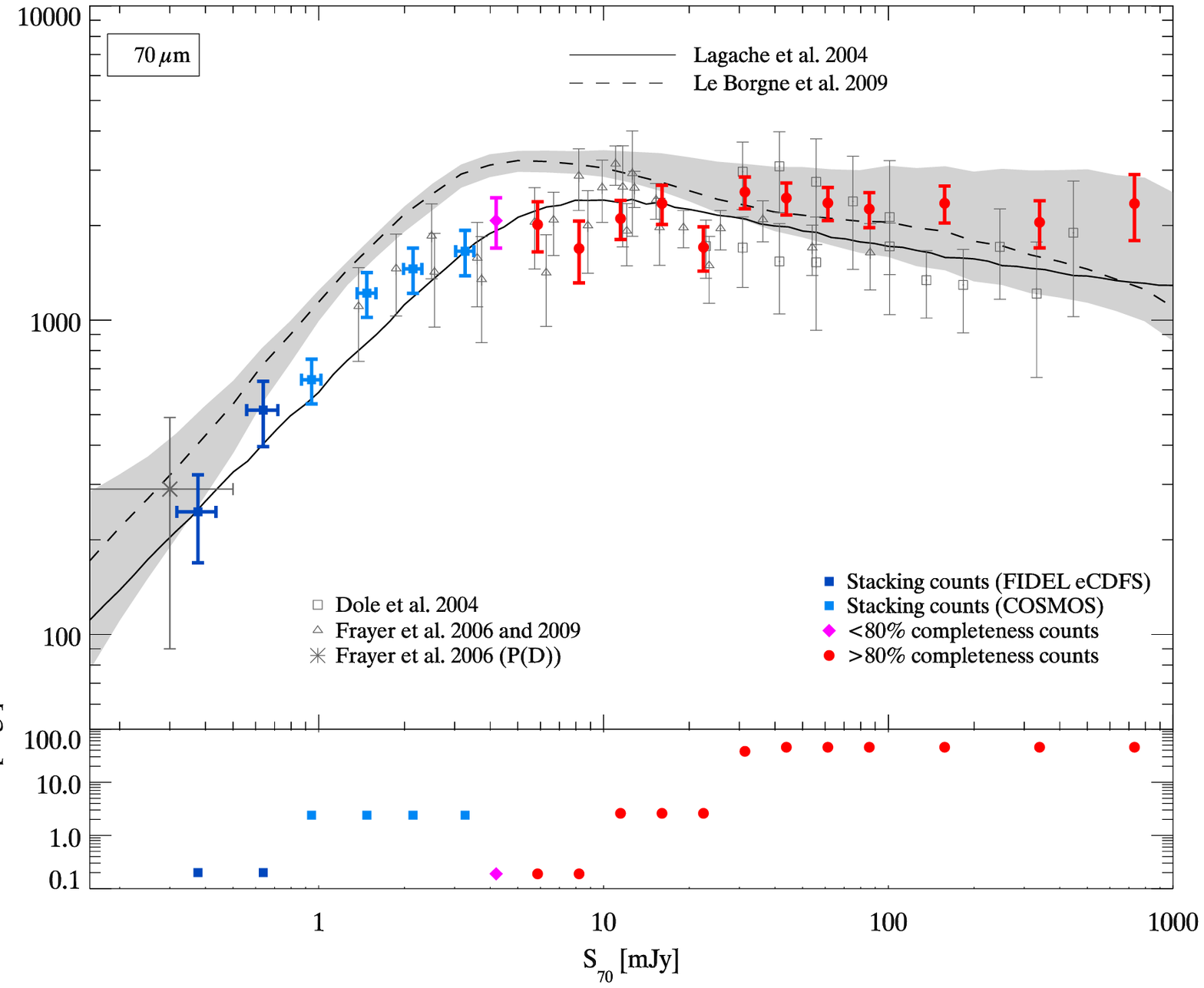}
\caption{Differential number counts at 70 $\mu$m. \textit{ Filled circle}: points obtained with $\ge$ 80\% completeness; \textit{filled diamond}: points obtained with less than 50\% spurious sources and less than 80\% completeness; \textit{filled square}: stacking number counts (clear: FIDEL eCDFS, dark: COSMOS); \textit{open square}: \citet{Dole2004} number counts in CDFS, Bootes and Marano; \textit{open triangle}: \citet{Frayer2006b} in GOODS and \citet{Frayer2009} in COSMOS; \textit{cross}: \cite{Frayer2006b} deduced from background fluctuations; \textit{ continuous line}: \citet{Lagache2004} model; \textit{dashed line and grey region}: \citet{Le_Borgne2009} model and 90\% confidence region. Error bars take into account clustering (see Sect. \ref{subsection:sectclus}) and calibration uncertainties \citep{Gordon2007}.}
\label{fig:70nc}
\end{figure*}

Counts in the flux density bins brighter than the 80\% completeness limit were obtained in the same way as at 24 $\mu$m (Fig. \ref{fig:70nc} and Table \ref{tab:70chart}). In addition, they were corrected from the Eddington bias (c.f. Sect. \ref{subsection:edd_bias}). We reached about 4.9 mJy at 80\% completeness (4.9 to 6.8 bin). We used CDFS below 80\% completeness limit to probe fainter flux density level. We cut these counts at 3.5 mJy. At this flux density, the spurious rate reached 50\%. We used a stacking analysis to probe fainter flux density levels (c.f. section \ref{section:stacksection}).\\

We can see breaks in the counts around 10~mJy and 20~mJy. These breaks appear between points built with a different set of fields. Our counts agree with earlier works of \citet{Dole2004}, \citet{Frayer2006b} and \citet{Frayer2009}. However, these works suppose only a Poissonian uncertainty, which underestimates the error bars (see Sect. \ref{subsection:sectclus}). Our data also agree well with these works. The \citet{Lagache2004} model agrees well with our data. The \citet{Le_Borgne2009} model gives a reasonable fit, despite an excess of about 30\% between 3~mJy and 10~mJy.\\

\subsection{160 $\mu$m number counts} 

\label{subsection:160normnc}

\begin{figure*}
\centering
\includegraphics{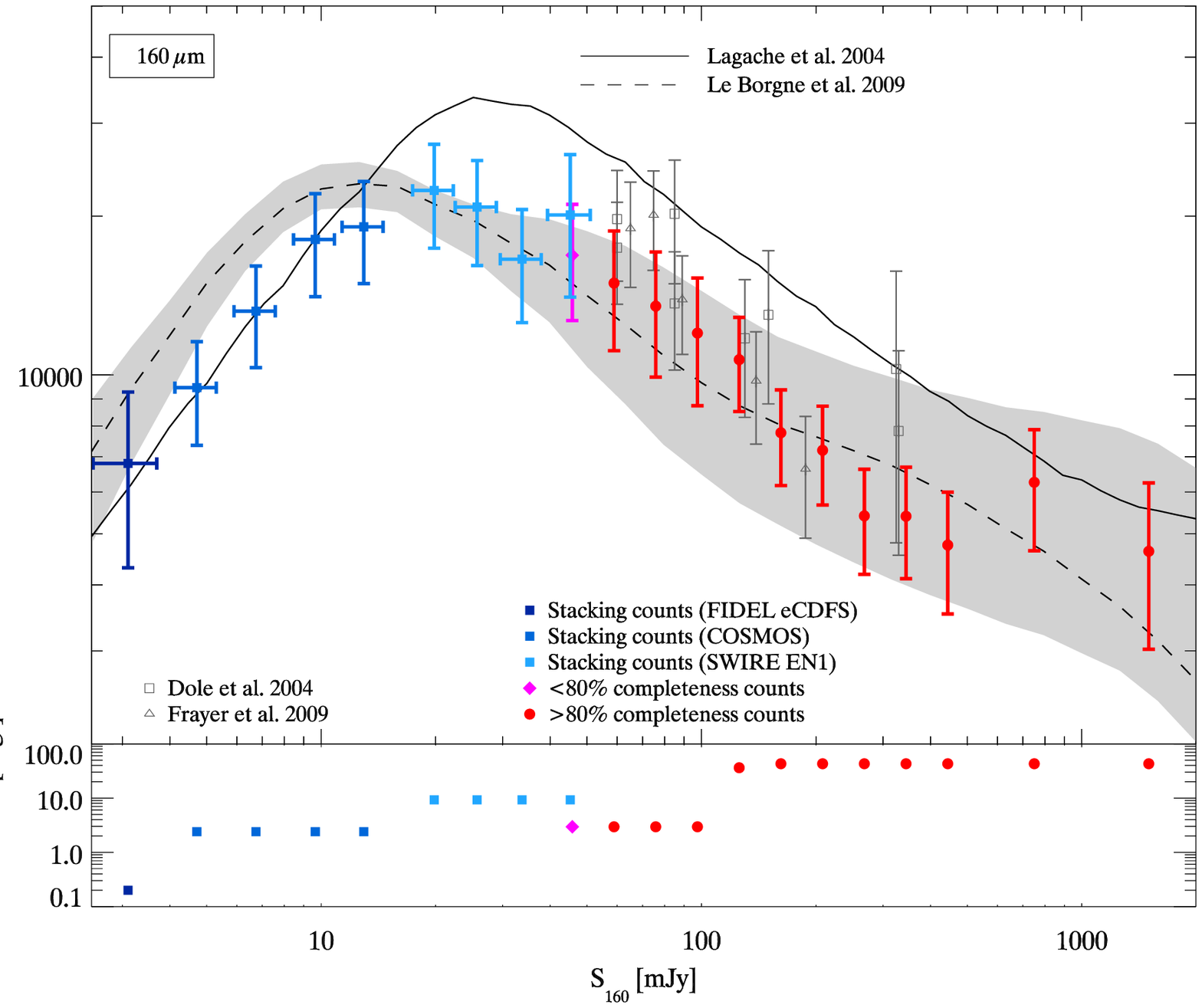}
\caption{Differential number counts at 160 $\mu$m. \textit{ Filled circle}: points obtained with $\ge$ 80\% completeness; \textit{filled diamond}: points obtained with less than 50\% spurious sources and less than 80\% completeness; \textit{filled square}: stacking number counts (clear: FIDEL/GTO CDFS, middle: COSMOS, dark: SWIRE EN1); \textit{open square}: \citet{Dole2004} number counts in CDFS and Marano; \textit{open triangle}: \citet{Frayer2009} in COSMOS; \textit{continuous line}: \citet{Lagache2004} model; \textit{dashed line and grey region}: \citet{Le_Borgne2009} model and 90\% confidence region. Error bars take into account clustering (see Sect. \ref{subsection:sectclus}) and calibration uncertainties \citep{Stansberry2007}.}
\label{fig:160nc}
\end{figure*}

The 160~$\mu$m number counts were obtained exactly in the same way as at 70 $\mu$m. We used COSMOS and EGS to probe counts below the 80\% completeness limit. We reached 51~mJy at  80\% completeness (51~mJy to 66~mJy bin) and 40~mJy for the 50\% spurious rate cut (figure \ref{fig:160nc} and table \ref{tab:160chart}). We used a stacking analysis to probe fainter flux density levels (c.f. Sect. \ref{section:stacksection}).\\

Our counts agree with the earlier works of \citet{Dole2004} and \citet{Frayer2009} . We find like \citet{Frayer2009} that the \citet{Lagache2004} model overestimates the counts by about 30\% above 50 mJy (see the discussion in Sect. \ref{subsection:160prob}). On the contrary, the \citet{Le_Borgne2009} model underpredicts the counts by about 20\% between 50~mJy and 150~mJy.\\

\subsection{Uncertainties on number counts including clustering}

\label{subsection:sectclus}

\citet{Shupe2008} showed that the SWIRE field-to-field variance is significantly higher than the Poisson noise (by a factor of three in some flux bins). They estimated their uncertainties on number counts with a field bootstrap method. We used a more formal method to deal with this problem.\\

The uncertainties on the number counts are Poissonian only if sources are distributed uniformly. But, actually, the infrared galaxies are clustered. The uncertainties must thus be computed taking into account clustering. We first measured the source clustering as a function of the flux density with the counts-in-cells moments (c-in-c) method \citep{Peebles1980,Szapudi1998,Blake2002}. We then computed the uncertainties knowing these clustering properties of the sources, the source density in the flux density bins and the field shapes. The details are explained in the appendix \ref{section:unc_clustering}.\\

This statistical uncertainty can be combined with the \textit{Spitzer} calibration uncertainty \citep{Engelbracht2007,Gordon2007,Stansberry2007} to compute the total uncertainty on differential number counts.\\

\section{Deeper FIR number counts using a stacking analysis}
\label{section:stacksection}

\subsection{Method}

The number counts  derived in Sect. \ref{section:counts} show that down to the 80\% completeness limit, the source surface density is 24100~deg$^{-2}$, 1200~deg$^{-2}$, and 220~deg$^{-2}$ at 24, 70, and 160~$\mu$m, respectively, i.e. 20 times (resp. 110 times) higher at 24 $\mu$m than at 70~$\mu$m (resp. 160~$\mu$m). These differences can be explained by the angular resolution decreasing with increasing wavelength, thus increasing confusion, and the noise properties of the detectors. There are thus many 24 $\mu$m sources without detected FIR counterparts. If we want to probe deeper into the FIR number counts, we can take advantage of the information provided by the 24 $\mu$m data, namely the existence of infrared galaxies not necessarily detected in the FIR, and their positions.\\

We used a stacking analysis \citep{Dole2006} to determine the FIR/MIR color as a function of the MIR flux. With this information, we can convert MIR counts into FIR counts.  The stacking technique consists in piling up very faint far-infrared galaxies which are not detected individually, but are detected at 24 $\mu$m. For this purpose, it makes use of the 24 $\mu$m data prior to tracking their undetected counterpart at 70~$\mu$m and 160~$\mu$m, where most of the bolometric luminosity arises. This method was used by \citet{Dole2006}, who managed to resolve the FIR CIB using 24 $\mu$m sources positions, as well many other authors (e.g \citet{Serjeant2004, Dye2006, Wang2006,Devlin2009,Dye2009, Marsden2009, Pascale2009}).\\

To derive the 70~$\mu$m or 160~$\mu$m versus 24 $\mu$m color, we stacked the FIR maps (cleaned of bright sources) at the positions of the 24 $\mu$m sources sorted by flux, and performed aperture photometry (same parameters as in Sect. \ref{section:se}). We thus get
\begin{equation}
 \overline{S_{FIR}} = f(\overline{S_{24}}),
 \label{eq:relcoul}
\end{equation}
where $\overline{S_{FIR}}$ is the average flux density in the FIR of the population selected at 24 $\mu$m, $\overline{S_{24}}$ the average flux density at 24 microns, and f is the function linking both quantities. We derive f empirically using the S$_{FIR}$ versus S$_{24}$ relation obtained from stacking.\\

 We checked that $f$ is a smooth monotonic function, in agreement with the expectation that the color varies smoothly with the redshift and the galaxy emission properties. Assuming that the individual sources follow this relation exactly, the FIR number counts could be deduced from
\begin{equation}
\frac{dN}{dS_{FIR}}\biggl | _{S_{FIR} = f(S_{24})} = \frac{dN}{dS_{24}}\biggl | _{S_{24}} \big /   \frac{dS_{FIR}}{dS_{24}}\biggl | _{S_{24}}.
\end{equation}
In practice, the two first terms are discrete. In addition, the last term is computed numerically in the same $S_{24}$ bin, using the two neighboring flux density bins (k-1 and k+1). We finally get
\begin{equation}
\frac{dN}{dS_{FIR}}(\overline{S_{FIR}}_k) = \left ( \frac{dN}{dS_{24}} \right )_{k}  \big /  \frac{\overline{S_{FIR,k+1}} - \overline{S_{FIR,k-1}}}{\overline{S_{24,k+1}} - \overline{S_{24,k-1}}},
\label{eq:stackcountsbin}
\end{equation}
where $<S_{FIR}>$ is measured by stacking.
In reality, sources do not follow Eq. \ref{eq:relcoul} exactly, but exhibit a scatter around this mean relation
\begin{equation}
 S_{FIR} = f(S_{24})+\sigma.
\end{equation}
Our method is still valid under the condition $\frac{\sigma}{f} \ll 1$, and we verify its validity with simulations (see next section).\\
 
 To obtain a better signal to noise ratio, we cleaned the resolved bright sources from the FIR maps prior to stacking. We used 8~$S_{24}$ bins per decade. We stacked a source only if the coverage was more than half of the median coverage of the map. Uncertainties on the FIR mean flux were estimated with a bootstrap method. Furthermore, knowing the uncertainties on the 24 $\mu$m number counts and the mean $S_{24}$ fluxes, we deduced the uncertainties on the FIR number counts according to Eq. \ref{eq:stackcountsbin}.\\ 
 
 At 70 $\mu$m, we used the FIDEL eCDFS (cleaned at  $S_{70}>10$~mJy) and the COSMOS (cleaned at  $S_{70}>50$~mJy) fields. At 160 $\mu$m, we used 160 $\mu$m the GTO CDFS (cleaned at  $S_{160}>60$~mJy), the COSMOS (cleaned at  $S_{160}>100$~mJy) and the SWIRE EN1 (no clean to probe the $S_{160}>20$~mJy sources) fields.\\
 
\subsection{Validation on simulations}
\label{subsection:valid}

\begin{figure}
\centering
\includegraphics{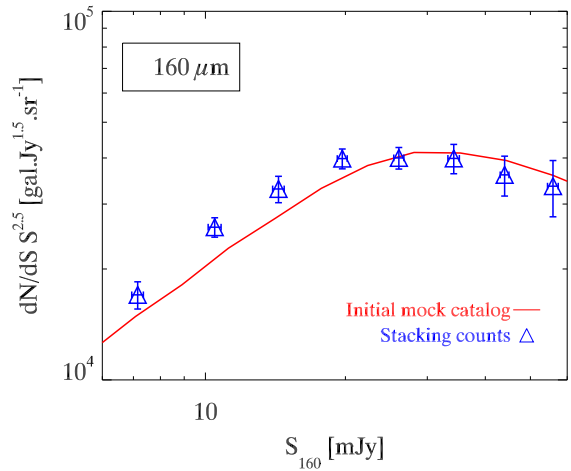}
\caption{\label{fig:valid_simu} Simulated number counts at 160 $\mu$m, computed from stacking counts \textit{(triangle)} and from the input mock catalog \textit{(solid line)}. The good agreement (better than 15\%) validates the stacking counts method (see Sect. \ref{subsection:valid}). Twenty realizations of the 2.9 deg$^2$ field maps with about 870~000 mock sources each were used.}
\end{figure}

We used the \citet{Fernandez-Conde2008} simulations\footnote{Publicly available at http://www.ias.u-psud.fr/irgalaxies/} to validate our method. These simulations are based on the \citet{Lagache2004} model, and include galaxy clustering. We employed 20 simulated mock catalogs of a 2.9 $deg^2$ field each, containing about 870~000 mock sources each. The simulated maps have the same pixel size as the actual Spitzer mosaics (1.2'', 4'', and 8'' at 24~$\mu$m, 70~$\mu$m, and 160~$\mu$m, resp.) and are convolved with our empirical PSF. A constant standard deviation Gaussian noise was added. We applied the same method as for the real data to produce stacking number counts. At 160 $\mu$m, for bins below 15~mJy, we cleaned the sources brighter than 50~mJy. Figure \ref{fig:valid_simu} shows 160 $\mu$m number counts from the mock catalogs down to $S_{160}=1$~mJy (diamond) and the number counts deduced from the stacking analysis described in the previous section (triangle). The error bars on the figure are the standard deviations of the 20 realization. There is a good agreement between the stacking counts and the classical counts (better than 15\%). Nevertheless, we observed a systematic bias, intrinsic to the method, of about 10\% in some flux density bins. We thus combined this 10\% error with the statistical uncertainties to compute our error bars. We also validated the estimation of the statistical uncertainty in the stacking counts: we check that the dispersion of the counts obtained by stacking, coming from different realizations, was compatible with our estimation of statistical uncertainties. The results are the same at 70 $\mu$m.\\

\subsection{Results}

\label{subsection:firstack}

At 70 $\mu$m, the stacking number counts reach 0.38 mJy (see Fig. \ref{fig:70nc} and Table \ref{tab:stack70}). The last stacking point is compatible with the \citet{Frayer2006b} P(D) constraint. The stacking points also agree very well with the \citet{Lagache2004} model. The \citet{Le_Borgne2009} model predicts slightly too many sources in 0.3 to 3 mJy range. The turnover around 3 mJy and the power law behavior of the faint counts ($S_{70} < 2 mJy$), observed by \citet{Frayer2006b} are confirmed with a better accuracy.\\

At 160 $\mu$m, the stacking counts reach 3.1~mJy (see Fig. \ref{fig:160nc} and Table \ref{tab:stack160}). We observed for the first time a turnover at about 20~mJy, and a power-law decrease at smaller flux densities. The stacking counts are lower than the \citet{Lagache2004} model around 20 mJy (about 30\%). Below 15 mJy, the stacking counts agree with this model. The \citet{Le_Borgne2009} model agree quite well with our points below 20 mJy. The results at 160 $\mu$m will be discussed in Sect. \ref{subsection:160prob}.\\

\begin{table}
\centering
\begin{tabular}{|l|r|rr|r|}
\hline
$<S>$ & $dN/dS.S^{2.5}$ & $\sigma_{clus.}$ & $\sigma_{clus.+calib.}$ & Field\\
\cline{1-4}
(in mJy) & \multicolumn{3}{|c|}{(in gal.Jy$^{1.5}$.sr$^{-1}$)} & \\
\hline
$    0.38\pm    0.05$ &   246. &    72. &    76. & FIDEL CDFS\\
$    0.64\pm    0.07$ &   517. &   109. &   122. & FIDEL CDFS\\
$    0.94\pm    0.03$ &   646. &    80. &   105. & COSMOS\\
$    1.48\pm    0.04$ &  1218. &   151. &   198. & COSMOS\\
$    2.14\pm    0.05$ &  1456. &   183. &   239. & COSMOS\\
$    3.27\pm    0.07$ &  1657. &   211. &   273. & COSMOS\\
\hline
\end{tabular}
\caption{\label{tab:stack70} Stacking extragalactic number counts at 70 $\mu$m. $\sigma_{clus.}$ is the uncertainty taking into account clustering (see Sect. \ref{subsection:sectclus}). $\sigma_{clus.+calib.}$ takes into account both clustering and calibration \citep{Gordon2007}.}

\begin{tabular}{|l|r|rr|r|}
\hline
$<S>$ & $dN/dS.S^{2.5}$ & $\sigma_{clus.}$ & $\sigma_{clus.+calib.}$ & Field\\
\cline{1-4}
(in mJy) & \multicolumn{3}{|c|}{(in gal.Jy$^{1.5}$.sr$^{-1}$)} & \\
\hline
$    3.11\pm    0.46$ &  6795. &  2163. &  2485. & GTO CDFS\\
$    4.71\pm    0.16$ &  9458. &  1236. &  2104. & COSMOS\\
$    6.74\pm    0.22$ & 13203. &  1627. &  2880. & COSMOS\\
$    9.65\pm    0.26$ & 18057. &  2307. &  3986. & COSMOS\\
$   12.95\pm    0.37$ & 19075. &  2388. &  4182. & COSMOS\\
$   19.82\pm    0.48$ & 22366. &  2944. &  4987. & SWIRE EN1\\
$   25.71\pm    0.81$ & 20798. &  2811. &  4682. & SWIRE EN1\\
$   33.74\pm    0.98$ & 16567. &  2671. &  4004. & SWIRE EN1\\
$   45.18\pm    2.08$ & 20089. &  4849. &  6049. & SWIRE EN1\\
\hline
\end{tabular}
\caption{\label{tab:stack160} Stacking extragalactic number counts at 160 $\mu$m. $\sigma_{clus.}$ is the uncertainty taking into account clustering (see Sect. \ref{subsection:sectclus}). $\sigma_{clus.+calib.}$ takes into account both clustering and calibration \citep{Stansberry2007}.}
\end{table}

\section{New lower limits and estimates of the CIB at 24~$\mu$m, 70~$\mu$m and 160~$\mu$m}

\subsection{24 $\mu$m CIB: lower limit and estimate}

\label{subsection:CIB24}

\begin{figure}
\centering
\includegraphics{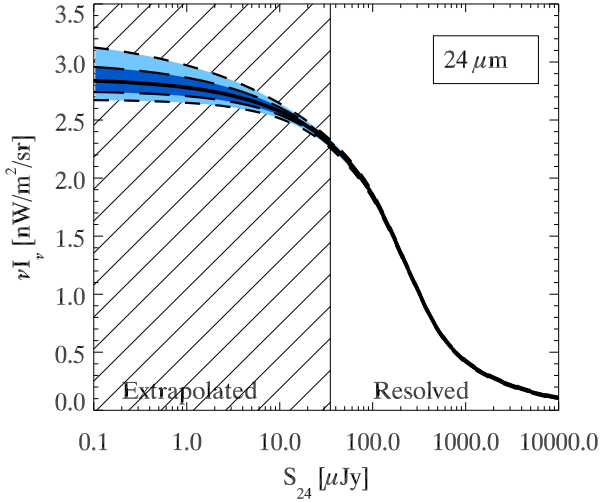}
\caption{\label{fig:int24} Cumulative contribution to the surface brightness of the 24 $\mu$m CIB as a function of $S_{24 \mu m}$. The colored area represents the 68\% and 95\% confidence level. The shaded area represents the $S_{24} < 35 \mu$Jy power-law extrapolation zone (see Sect. \ref{subsection:CIB24}). The 4\% calibration uncertainty is not represented. The table corresponding to this figure is available online at http://www.ias.u-psud.fr/irgalaxies/.}
\end{figure}

By integrating the measured 24~$\mu$m number counts between 35~$\mu$Jy and 0.1 Jy, we can estimate a lower value of the CIB at this wavelength. The counts were integrated with a trapeze method. We estimated the uncertainty on the integral by adding (on 10000 realisations) a random Gaussian error to each data point with the $\sigma$ given by the count uncertainties taking into account clustering. We then added the 4\% calibration error of the instrument \citep{Engelbracht2007}. We found $2.26_{-0.09}^{+0.09}$ nW.m$^{-2}$.sr$^{-1}$. The very bright source counts ($S_{24} > 0.1$ Jy) are supposed to be euclidian ($dN/dS = C_{eucl} S^{2.5}$). We used the three brightest points to estimate $C_{eucl}$. We found a contribution to the CIB of $0.032_{-0.003}^{+0.003}$~nW.m$^{-2}$.sr$^{-1}$. Consequently, very bright sources extrapolation is not critical for the CIB estimation (1\% of CIB). The contribution of $S_{24} >$~35~$\mu$Jy is thus $2.29_{-0.09}^{+0.09}$ nW.m$^{-2}$.sr$^{-1}$ (c.f. Table \ref{tab:CIBchart}).\\

We might have wanted to estimate the CIB value at 24 $\mu$m. To do so, we needed to extrapolate the number counts on the faint end. Below 100 $\mu$Jy, the number counts exhibit a power-law behavior (Fig. \ref{fig:24nc}). We assumed that this behavior (of the form $dN/dS = C_{faint} S^{r}$) still holds below 35 $\mu$Jy. r and $C_{faint}$ are determined using the four faintest bins. We found $r = 1.45 \pm 0.10$ (compatible with $1.5 \pm 0.1$ of \citet{Papovitch2004}). Our new estimate of the CIB at 24 $\mu$m due to infrared galaxies is thus $2.86_{-0.16}^{+0.19}$ nW.m$^{-2}$.sr$^{-1}$. The results are plotted in Fig. \ref{fig:int24}. We conclude that resolved sources down to $S_{24}=35 \mu$Jy account for 80\% of the CIB at this wavelength.\\

\subsection{70 ~$\mu$m and 160~$\mu$m CIB: lower limit and estimate}

\label{subsection:CIBFIR}

\begin{figure}
\centering
\includegraphics{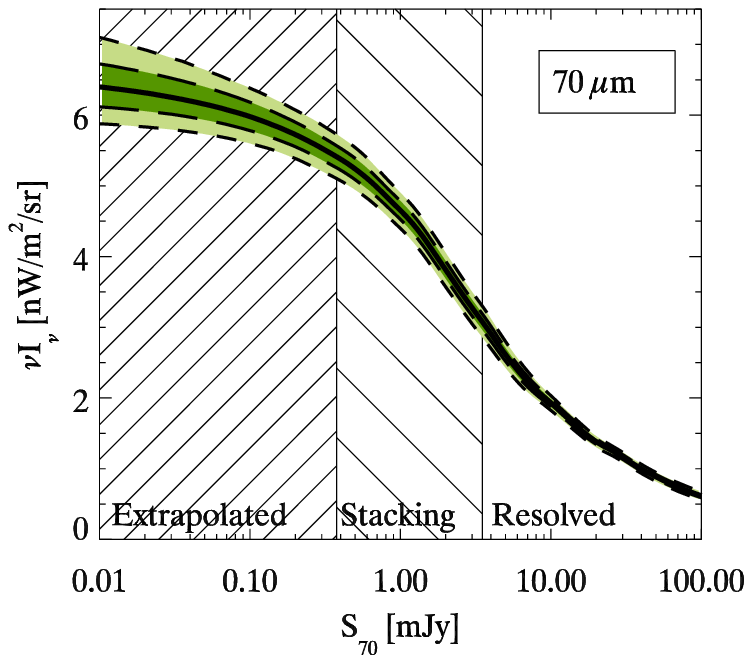}
\caption{\label{fig:int70} Cumulative contribution to the surface brightness of the 70 $\mu$m CIB as a function of $S_{70 \mu m}$. The colored area represents the 68\% and 95\% confidence level. The shaded areas represent the $0.38 < S_{70} < 3.3$ mJy stacking counts zone and the $S_{70} < 0.38$ mJy power-law extrapolation zone (see Sect. \ref{subsection:CIBFIR}). The 7\% calibration uncertainty is not represented. The table corresponding to this figure is available online at http://www.ias.u-psud.fr/irgalaxies/.}
\includegraphics{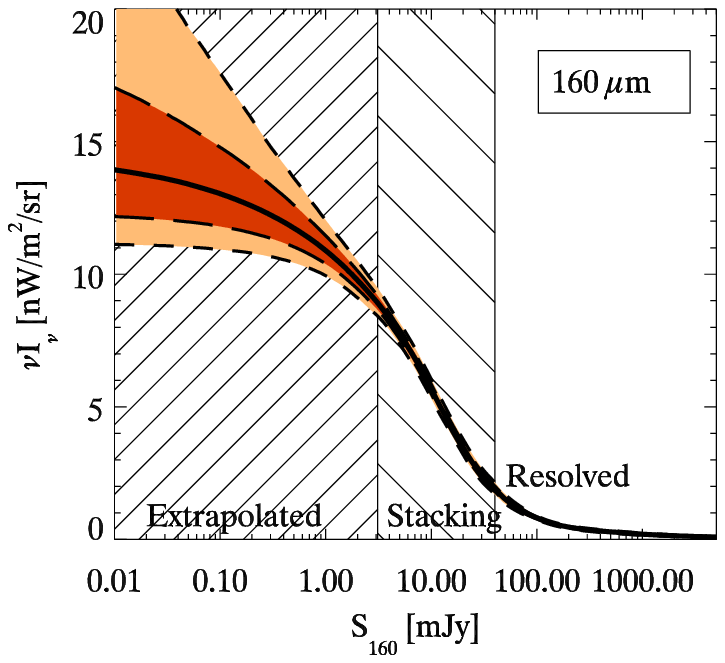}
\caption{\label{fig:int160} Cumulative contribution to the surface brightness of the 160 $\mu$m CIB as a function of $S_{160 \mu m}$. The colored area represents the 68\% and 95\% confidence level. The shaded areas represent the $3.1 < S_{160} < 45$ mJy stacking counts zone and the $S_{160} < 3.1$ mJy power-law extrapolation zone (see Sect. \ref{subsection:CIBFIR}). The 12\% calibration uncertainty is not represented. The table corresponding to this figure is available online at http://www.ias.u-psud.fr/irgalaxies/.}
\end{figure}

At 70~$\mu$m and 160~$\mu$m, the integration of the number counts was done in the same way as at 24 $\mu$m, except for the stacking counts, which are correlated. To compute the uncertainties on the integral, we added (on 10000 realizations) a Gaussian error simultaneously to the three quantities and completely recomputed the associated stacking counts: 1- the mean density flux given by the stacking; 2- the 24~$\mu$m number counts; 3- the mean 24 $\mu$m flux density. 
At 70~$\mu$m, and 160~$\mu$m, the calibration uncertainty is 7 \% \citep{Gordon2007} and 12\% \citep{Stansberry2007}, respectively. We estimated the CIB surface brightness contribution of resolved sources ($S_{70} >$~3.5~mJy and $S_{160} >$~40~mJy) of $3.1_{-0.2}^{+0.2}$~nW.m$^{-2}$.sr$^{-1}$ and $1.0_{-0.1}^{+0.1}$~nW.m$^{-2}$.sr$^{-1}$. The contribution of $S_{70}>0.38$~mJy and $S_{160}>3.1$~mJy is $5.4_{-0.4}^{+0.4}$~nW.m$^{-2}$.sr$^{-1}$ and $8.9_{-1.1}^{+1.1}$~nW.m$^{-2}$.sr$^{-1}$, respectively.\\
    
Below 2~mJy at 70~$\mu$m, and 10~mJy at 160~$\mu$m, the stacking counts are compatible with a power-law. Like at 24 $\mu$m, we assumed that this behavior can be extrapolated and determined the law with the five faintest bins at 70~$\mu$m, and the four faintest at 160 $\mu$m. We found a slope $r = 1.50 \pm 0.14$ at 70~$\mu$m, and $1.61 \pm 0.21$ at 160~$\mu$m. The slope of the number counts at 70~$\mu$m is compatible with the \citet{Frayer2006b} value ($1.63 \pm 0.34$). The slope at 160~$\mu$m is measured for the first time. Our new estimate of the CIB at 70~$\mu$m and 160~$\mu$m due to infrared galaxies is thus $6.6_{-0.6}^{+0.7}$~nW.m$^{-2}$.sr$^{-1}$, and $14.6_{-2.9}^{+7.1}$~nW.m$^{-2}$.sr$^{-1}$, respectively. We conclude that resolved and stacking-studied populations account for 82\% and 62\% of the CIB at 70~$\mu$m and 160~$\mu$m, respectively. These results are summarized in Table \ref{tab:CIBchart}, and Fig. \ref{fig:int70} and \ref{fig:int160}.\\

\begin{table}
\centering
\begin{tabular}{|l|l|r|r|r|}
\cline{3-5}
\multicolumn{2}{c|}{ } & 24 $\mu$m & 70 $\mu$m & 160 $\mu$m \\
\hline
$S_{cut,resolved}$ & mJy & 0.035 & 3.50 & 40.0\\
$S_{cut,stacking}$ & & - & 0.38 &  3.1\\
\hline
$\nu B_{\nu,resolved}$ & nW.m$^{-2}$.sr$^{-1}$ & $2.29_{-0.09}^{+0.09}$ & $ 3.1_{- 0.2}^{+ 0.2}$ & $ 1.0_{- 0.1}^{+ 0.1}$\\
$\nu B_{\nu,resolved+stacking}$ & & - & $ 5.4_{- 0.4}^{+ 0.4}$ & $ 8.9_{- 1.1}^{+ 1.1}$ \\
$\nu B_{\nu,tot}$ & & $2.86_{-0.16}^{+0.19}$ & $ 6.6_{- 0.6}^{+ 0.7}$ & $14.6_{- 2.9}^{+ 7.1}$\\
\hline
\end{tabular}
\caption{\label{tab:CIBchart} Summary of CIB results found in this article.}
\end{table}

\section{Discussion}

\subsection{New lower limits of the CIB}

\label{CIBdisc}

The estimations of CIB based on number counts ignore a potential diffuse infrared emission like dust in galaxy clusters \citep{Montier2005}. The extrapolation of the faint source counts supposes no low luminosity population, like population III stars or faint unseen galaxies. Accordingly, this type of measurement can provide in principle only a lower limit.\\

At 24 $\mu$m, \citet{Papovitch2004} found $2.7_{1.1}^{-0.7}$~nW.m$^{-2}$.sr$^{-1}$ using the counts and the extrapolation of the faint source counts. We agree with this work and significantly reduced the uncertainties on this estimation. \citet{Dole2006} found a contribution of $1.93 \pm 0.23$~nW.m$^{-2}$.sr$^{-1}$ for $S_{24} > 60 \mu$Jy sources (after dividing their results by 1.12 to correct an aperture error in their photometry at 24 $\mu$m). Our analysis gives  $2.10 \pm 0.08$~nW.m$^{-2}$.sr$^{-1}$ for a cut at 60 ~$\mu$Jy, which agrees very well. \citet{Rodighiero2006} gave a total value of $2.6$~nW.m$^{-2}$.sr$^{-1}$,  without any error bar. \citet{Chary2004}  found $2.0 \pm 0.2 $~nW.m$^{-2}$.sr$^{-1}$, by integrating sources between 20 and 1000~$\mu$Jy (we find $2.02 \pm 0.10$ for the same interval).\\

At 70 $\mu$m, using the number counts in the ultra deep GOODS-N and a P(D) analysis, \citet{Frayer2006b} found a $S_{70} >$~0.3 mJy source contribution to the 70 $\mu$m CIB of $5.5 \pm 1.1$~nW.m$^{-2}$.sr$^{-1}$. Using the stacking counts, we found $5.5 \pm 0.4$~nW.m$^{-2}$.sr$^{-1}$ for the same cut, in excellent agreement and with improved uncertainties. In \citet{Dole2006}, the contribution at 70 $\mu$m of the $S_{24} <$~60~$\mu$Jy sources was computed with an extrapolation of the 24 $\mu$m number counts and the 70/24 color. They found $7.1 \pm 1.0$~nW.m$^{-2}$.sr$^{-1}$, but the uncertainty on the extrapolation took only into account the uncertainty on the 70/24 color and not the uncertainty on the extrapolated 24 $\mu$m contribution, and was thus slightly underestimated. This is in agrees with our estimation.\\

At 160 $\mu$m they found with the same method, $17.4 \pm 2.1$~nW.m$^{-2}$.sr$^{-1}$ (a corrective factor of 1.3 was applied due to an error on the map pixel size). This estimation is a little bit higher than our estimation, and can be explained by a small contribution (of the order of 15\%) of the source clustering \citep{Bavouzet_thesis}.\\

Our results can also be compared with direct measurements made by absolute photometers.  These methods are biased by the foreground modeling, but do not ignore the extended emission. \citet{Fixsen1998} found a CIB brightness of 13.7$\pm$3.0 at 160~$\mu$m, in excellent agreement with our estimation ($14.6_{-2.9}^{+7.1}$~nW.m$^{-2}$). From the discussion in \citet{Dole2006} (Sect. 4.1), the \citet{Lagache2000} DIRBE WHAM (FIRAS calibration) estimation at 140~$\mu$m and 240~$\mu$m of 12~nW.m$^{-2}$.sr$^{-1}$ and 12.2~nW.m$^{-2}$.sr$^{-1}$ can be also compared with our value at 160 $\mu$m. A more recent work of \citet{Odegard2007} found 25.0$\pm$6.9 and 13.6$\pm$2.5~nW.m$^{-2}$.sr$^{-1}$ at 140~$\mu$m and 240~$\mu$m respectively (resp. 15$\pm$5.9~nW.m$^{-2}$.sr$^{-1}$ and 12.7$\pm$1.6~nW.m$^{-2}$.sr$^{-1}$ with the FIRAS scale). Using ISOPHOT data, \citet{Juvela2009} give an estimation of the CIB surface brightness between 150$\mu$m and 180 $\mu$m of 20.25$\pm$6.0$\pm$5.6~nW.m$^{-2}$.sr$^{-1}$.\\

The total brightness due to infrared galaxies at 160~$\mu$m corresponds to the total CIB level at this wavelength. We thus have probably resolved the CIB at this wavelength. Nevertheless, the uncertainties are relatively large, and other minor CIB contributors cannot be excluded.\\

In addition, upper limits can be deduced indirectly from blazar high energy spectrum. \citet{Stecker1997} gave an upper limit of 4~nW.m$^{-2}$.s$r^{-1}$ at 20 $\mu$m using Mkn 421. \citet{Renault2001} found an upper limit of 4.7~ nW.m$^{-2}$.sr$^{-1}$ between 5 and 15 $\mu$m with Mkn 501. This is consistent with our lower limit at 24 $\mu$m. \\

\begin{figure*}
\centering
\includegraphics{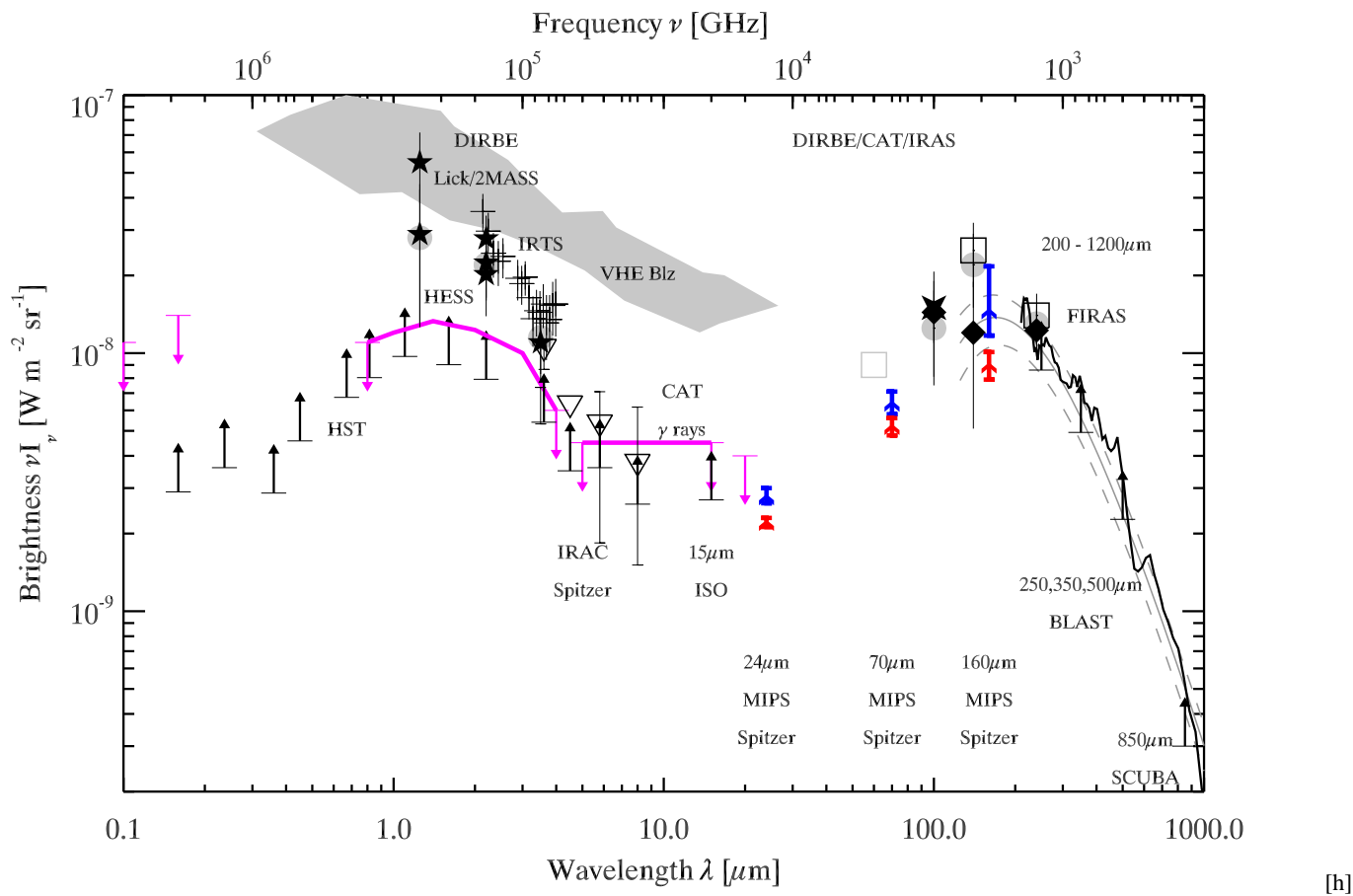}[h]
\caption{Current measurement of the extragalactic background light spectral energy distribution from 100 nm to 1mm, with the cosmic optical background (COB, $\lambda < 8 \, \mu$m) and cosmic infrared background (CIB, $\lambda > 8 \, \mu$m). Our new points at 24~$\mu$m, 70~$\mu$m and 160~$\mu$m are plotted (\textit{triangle}). Lower (red) triangles correspond to the CIB resolved with the number counts and stacking counts. Upper (blue) triangles correspond to the total extrapolated CIB due to infrared galaxies. BLAST lower limits at 250~$\mu$m, 350~$\mu$m and 500~$\mu$m \citep{Devlin2009,Marsden2009} are represented in \textit{black arrows}. The FIRAS measurements of \citet{Fixsen1998} between 125~$\mu$m and 2000~$\mu$m are plotted with a \textit{grey solid line}, and the 1-$\sigma$ confidence region with a \textit{grey dashed line}. Other points come from different authors (see \citet{Dole2006} for complete details). Old MIPS points are not plotted for clarity.}
\label{fig:ebl}
\end{figure*}

An update of the synthetic EBL SED of \citet{Dole2006} with the new BLAST (balloon-borne large-aperture submillimeter telescope) lower limits from \citet{Devlin2009} and our values is plotted in Fig. \ref{fig:ebl}. The  BLAST lower limits are obtained by stacking of the Spitzer 24~$\mu$m sources at  250~$\mu$m, 350~$\mu$m and 500~$\mu$m \citep{Devlin2009,Marsden2009}.\\ 

\subsection{160 $\mu$m number counts}

\label{subsection:160prob}

At most, we observed a 30\% overestimation of the \citet{Lagache2004} model compared to the 160 $\mu$m number counts (Sect. \ref{subsection:160normnc} and \ref{subsection:firstack} and Fig. \ref{fig:160nc}), despite good fits at other wavelengths. This model uses mean SEDs of galaxies sorted into two populations (starburst and cold), whose luminosity functions evolve separately with the redshift. A possible explanation of the model excess is a slightly too high density of local cold galaxies. By decreasing the density of this local population a little, the model might be able to better fit the 160 $\mu$m number counts without significantly affecting other wavelengths, especially at 70 $\mu$m (more sensitive to warm dust rather than cold dust), and in the submillimetre range (more sensitive to redshifted cold dust at faint flux densities for wavelengths larger than 500 $\mu$m).\\ 

The \citet{Le_Borgne2009} model slightly overpredicts faint 160 $\mu$m sources, probably because of the presence of too many galaxies at high redshift; this trend is also seen at 70 $\mu$m. With our number counts as new constraints, their inversion should give more accurate parameters.\\

\textit{Herschel} was successfully launched on May 14th, 2009 (together with \textit{Planck}). It will observe infrared galaxies between 70~$\mu$m and 500~$\mu$m with an improved sensitivity. It will be possible to directly observe the cold dust spectrum of high-z ULIRG (ultra luminous infrared galaxy) and medium-z LIRG (luminous infrared galaxy). PACS (Photodetectors Array Camera and Spectrometer) will make photometric surveys in three bands centred on 70~$\mu$m, 100~$\mu$m and 160~$\mu$m. \textit{Herschel} will allow us to resolve a significant fraction of the background at these wavelengths \citep{Lagache2003,Le_Borgne2009}. SPIRE (spectral and photometric imaging receiver) will observe around 250~$\mu$m, 350~$\mu$m and 500 $\mu$m, and will be quickly confusion limited. In both cases, the stacking analysis will allow us to probe fainter flux density levels, as it is complementary to \textit{Spitzer} and BLAST.\\

\section{Conclusion}

With a large sample of public \textit{Spitzer} extragalactic maps, we built new deep, homogeneous, high-statistics number counts in three MIPS bands at 24~$\mu$m, 70~$\mu$m and 160~$\mu$m.\\

At 24 $\mu$m, the results agree with previous works. These counts are derived from the widest surface ever used at this wavelength (53.6 deg$^2$). Using these counts, we give an accurate estimation of the galaxy contribution to the CIB at this wavelength ($2.86_{-0.16}^{+0.19}$ nW.m$^{-2}$.sr$^{-1}$).\\

At 70 $\mu$m, we used the stacking method to determine the counts below the detection limit of individual sources, by reaching 0.38~mJy, allowing us to probe the faint flux density slope of differential number counts. With this information, we deduced the total contribution of galaxies to the CIB at this wavelength ($6.6_{-0.6}^{+0.7} nW.m^{-2}.sr^{-1}$).\\

At 160 $\mu$m, our counts reached 3~mJy with a stacking analysis. We exhibited for the first time the maximum in differential number counts around 20~mJy and the power-law behavior below 10~mJy. We deduced the total contribution of galaxies to the CIB at this wavelength ($14.6_{-2.9}^{+7.1} nW.m^{-2}.sr^{-1}$). \textit{Herschel} will likely probe flux densities down to about 10~mJy at this wavelength (confusion limit, \citet{Le_Borgne2009}).\\                                                                                                                                                                                                                                                                                                                                                                                                                                                                                                                                                                                                                                                                                                                                                                                                                                                                                                                                                                                                                                                                                                                                                                                                                                                                                                                                                                                                                                                                                                                                                                                                                                                                                                                                                                                                                                                                                                                                                                                                                                                                                                

The uncertainties on the number counts used in this work take carefully into account the galaxy clustering, which is measured with the "counts-in-cells" method.\\

We presented a method to build very deep number counts with the information provided by shorter wavelength data (MIPS 24 $\mu$m) and a stacking analysis. This tool could be used on \textit{Herschel} SPIRE data with a PACS prior to probe fainter flux densities in the submillimetre range.\\   

We publicly release on the website http://www.ias.u-psud.fr/irgal/, the following products: PSF, number counts and CIB contributions. We also release a stacking library software written in IDL.\\

\begin{acknowledgements}
We wish to acknowledge G. Lagache, who has generated simulations used in this work. We acknowledge J.L. Puget, G. Lagache, D. Marcillac, B. Bertincourt, A. Penin and all members of the cosmology group of IAS for their comments and suggestions. We wish to thank the members of ANR D-SIGALE for their valuable comments, in particular D. Le Borgne for providing us an electronic version of his model. We also thank E. Le Floc'h for quickly providing us the table of his counts. This work is based in part on archival data obtained with the Spitzer Space Telescope, which is operated by the Jet Propulsion Laboratory, California Institute of Technology under a contract with NASA. Support for this work was provided by an award issued by JPL/Caltech. This publication makes use of data products from the Two Micron All Sky Survey, which is a joint project of the University of Massachusetts and the Infrared Processing and Analysis Center/California Institute of Technology, funded by the National Aeronautics and Space Administration and the National Science Foundation.
\end{acknowledgements}

\bibliographystyle{aa}

\bibliography{biblio}

\begin{thebibliography}{60}
\expandafter\ifx\csname natexlab\endcsname\relax\def\natexlab#1{#1}\fi

\bibitem[{{Arendt} {et~al.}(1998){Arendt}, {Odegard}, {Weiland}, {Sodroski},
  {Hauser}, {Dwek}, {Kelsall}, {Moseley}, {Silverberg}, {Leisawitz},
  {Mitchell}, {Reach}, \& {Wright}}]{Arendt1998}
{Arendt}, R.~G., {Odegard}, N., {Weiland}, J.~L., {et~al.} 1998, \apj, 508, 74

\bibitem[{{Bavouzet}(2008)}]{Bavouzet_thesis}
{Bavouzet}, N. 2008, PhD thesis, Universit\'e Paris-Sud 11
  http://tel.archives-ouvertes.fr/tel-00363975/

\bibitem[{{Bertin} \& {Arnouts}(1996)}]{Bertin1996}
{Bertin}, E. \& {Arnouts}, S. 1996, \aaps, 117, 393

\bibitem[{{Blake} \& {Wall}(2002)}]{Blake2002}
{Blake}, C. \& {Wall}, J. 2002, \mnras, 337, 993

\bibitem[{{Chary} {et~al.}(2004){Chary}, {Casertano}, {Dickinson}, {Ferguson},
  {Eisenhardt}, {Elbaz}, {Grogin}, {Moustakas}, {Reach}, \& {Yan}}]{Chary2004}
{Chary}, R., {Casertano}, S., {Dickinson}, M.~E., {et~al.} 2004, \apjs, 154, 80

\bibitem[{{Devlin} {et~al.}(2009){Devlin}, {Ade}, {Aretxaga}, {Bock}, {Chapin},
  {Griffin}, {Gundersen}, {Halpern}, {Hargrave}, {Hughes}, {Klein}, {Marsden},
  {Martin}, {Mauskopf}, {Moncelsi}, {Netterfield}, {Ngo}, {Olmi}, {Pascale},
  {Patanchon}, {Rex}, {Scott}, {Semisch}, {Thomas}, {Truch}, {Tucker},
  {Tucker}, {Viero}, \& {Wiebe}}]{Devlin2009}
{Devlin}, M.~J., {Ade}, P.~A.~R., {Aretxaga}, I., {et~al.} 2009, \nat, 458, 737

\bibitem[{{Diolaiti} {et~al.}(2000){Diolaiti}, {Bendinelli}, {Bonaccini},
  {Close}, {Currie}, \& {Parmeggiani}}]{Diolaiti2000}
{Diolaiti}, E., {Bendinelli}, O., {Bonaccini}, D., {et~al.} 2000, \aaps, 147,
  335

\bibitem[{{Dole} {et~al.}(2001){Dole}, {Gispert}, {Lagache}, {Puget},
  {Bouchet}, {Cesarsky}, {Ciliegi}, {Clements}, {Dennefeld}, {D{\'e}sert},
  {Elbaz}, {Franceschini}, {Guiderdoni}, {Harwit}, {Lemke}, {Moorwood},
  {Oliver}, {Reach}, {Rowan-Robinson}, \& {Stickel}}]{Dole2001}
{Dole}, H., {Gispert}, R., {Lagache}, G., {et~al.} 2001, \aap, 372, 364

\bibitem[{{Dole} {et~al.}(2003){Dole}, {Lagache}, \& {Puget}}]{Dole2003}
{Dole}, H., {Lagache}, G., \& {Puget}, J.-L. 2003, \apj, 585, 617

\bibitem[{{Dole} {et~al.}(2006){Dole}, {Lagache}, {Puget}, {Caputi},
  {Fern{\'a}ndez-Conde}, {Le Floc'h}, {Papovich}, {P{\'e}rez-Gonz{\'a}lez},
  {Rieke}, \& {Blaylock}}]{Dole2006}
{Dole}, H., {Lagache}, G., {Puget}, J.-L., {et~al.} 2006, \aap, 451, 417

\bibitem[{{Dole} {et~al.}(2004){Dole}, {Le Floc'h}, {P{\'e}rez-Gonz{\'a}lez},
  {Papovich}, {Egami}, {Lagache}, {Alonso-Herrero}, {Engelbracht}, {Gordon},
  {Hines}, {Krause}, {Misselt}, {Morrison}, {Rieke}, {Rieke}, {Rigby}, {Young},
  {Bai}, {Blaylock}, {Neugebauer}, {Beichman}, {Frayer}, {Mould}, \&
  {Richards}}]{Dole2004}
{Dole}, H., {Le Floc'h}, E., {P{\'e}rez-Gonz{\'a}lez}, P.~G., {et~al.} 2004,
  \apjs, 154, 87

\bibitem[{{Dye} {et~al.}(2009){Dye}, {Ade}, {Bock}, {Chapin}, {Devlin},
  {Dunlop}, {Eales}, {Griffin}, {Gundersen}, {Halpern}, {Hargrave}, {Hughes},
  {Klein}, {Magnelli}, {Marsden}, {Mauskopf}, {Moncelsi}, {Netterfield},
  {Olmi}, {Pascale}, {Patanchon}, {Rex}, {Scott}, {Semisch}, {Targett},
  {Thomas}, {Truch}, {Tucker}, {Tucker}, {Viero}, \& {Wiebe}}]{Dye2009}
{Dye}, S., {Ade}, P.~A.~R., {Bock}, J.~J., {et~al.} 2009, \apj, 703, 285

\bibitem[{{Dye} {et~al.}(2006){Dye}, {Eales}, {Ashby}, {Huang}, {Webb},
  {Barmby}, {Lilly}, {Brodwin}, {McCracken}, {Egami}, \& {Fazio}}]{Dye2006}
{Dye}, S., {Eales}, S.~A., {Ashby}, M.~L.~N., {et~al.} 2006, \apj, 644, 769

\bibitem[{{Eddington}(1913)}]{Eddington1913}
{Eddington}, A.~S. 1913, \mnras, 73, 359

\bibitem[{{Eddington}(1940)}]{Eddington1940}
{Eddington}, Sir, A.~S. 1940, \mnras, 100, 354

\bibitem[{{Elbaz} {et~al.}(2002){Elbaz}, {Cesarsky}, {Chanial}, {Aussel},
  {Franceschini}, {Fadda}, \& {Chary}}]{Elbaz1999}
{Elbaz}, D., {Cesarsky}, C.~J., {Chanial}, P., {et~al.} 2002, \aap, 384, 848

\bibitem[{{Engelbracht} {et~al.}(2007){Engelbracht}, {Blaylock}, {Su}, {Rho},
  {Rieke}, {Muzerolle}, {Padgett}, {Hines}, {Gordon}, {Fadda},
  {Noriega-Crespo}, {Kelly}, {Latter}, {Hinz}, {Misselt}, {Morrison},
  {Stansberry}, {Shupe}, {Stolovy}, {Wheaton}, {Young}, {Neugebauer},
  {Wachter}, {P{\'e}rez-Gonz{\'a}lez}, {Frayer}, \&
  {Marleau}}]{Engelbracht2007}
{Engelbracht}, C.~W., {Blaylock}, M., {Su}, K.~Y.~L., {et~al.} 2007, \pasp,
  119, 994

\bibitem[{{Fernandez-Conde} {et~al.}(2008){Fernandez-Conde}, {Lagache},
  {Puget}, \& {Dole}}]{Fernandez-Conde2008}
{Fernandez-Conde}, N., {Lagache}, G., {Puget}, J.-L., \& {Dole}, H. 2008, \aap,
  481, 885

\bibitem[{{Fixsen} {et~al.}(1998){Fixsen}, {Dwek}, {Mather}, {Bennett}, \&
  {Shafer}}]{Fixsen1998}
{Fixsen}, D.~J., {Dwek}, E., {Mather}, J.~C., {Bennett}, C.~L., \& {Shafer},
  R.~A. 1998, \apj, 508, 123

\bibitem[{{Franceschini} {et~al.}(2009){Franceschini}, {Rodighiero}, {Vaccari},
  {Marchetti}, \& {Mainetti}}]{Franceschini2009}
{Franceschini}, A., {Rodighiero}, G., {Vaccari}, M., {Marchetti}, L., \&
  {Mainetti}, G. 2009, ArXiv e-prints

\bibitem[{{Frayer} {et~al.}(2006){Frayer}, {Huynh}, {Chary}, {Dickinson},
  {Elbaz}, {Fadda}, {Surace}, {Teplitz}, {Yan}, \& {Mobasher}}]{Frayer2006b}
{Frayer}, D.~T., {Huynh}, M.~T., {Chary}, R., {et~al.} 2006, \apjl, 647, L9

\bibitem[{{Frayer} {et~al.}(2009){Frayer}, {Sanders}, {Surace}, {Aussel},
  {Salvato}, {Le Floc'h}, {Huynh}, {Scoville}, {Afonso-Luis}, {Bhattacharya},
  {Capak}, {Fadda}, {Fu}, {Helou}, {Ilbert}, {Kartaltepe}, {Koekemoer}, {Lee},
  {Murphy}, {Sargent}, {Schinnerer}, {Sheth}, {Shopbell}, {Shupe}, \&
  {Yan}}]{Frayer2009}
{Frayer}, D.~T., {Sanders}, D.~B., {Surace}, J.~A., {et~al.} 2009, \aj, 138,
  1261

\bibitem[{{Gispert} {et~al.}(2000){Gispert}, {Lagache}, \&
  {Puget}}]{Gispert2000}
{Gispert}, R., {Lagache}, G., \& {Puget}, J.~L. 2000, \aap, 360, 1

\bibitem[{{Gordon} {et~al.}(2007){Gordon}, {Engelbracht}, {Fadda},
  {Stansberry}, {Wachter}, {Frayer}, {Rieke}, {Noriega-Crespo}, {Latter},
  {Young}, {Neugebauer}, {Balog}, {Beeman}, {Dole}, {Egami}, {Haller}, {Hines},
  {Kelly}, {Marleau}, {Misselt}, {Morrison}, {P{\'e}rez-Gonz{\'a}lez}, {Rho},
  \& {Wheaton}}]{Gordon2007}
{Gordon}, K.~D., {Engelbracht}, C.~W., {Fadda}, D., {et~al.} 2007, \pasp, 119,
  1019

\bibitem[{{Hauser} {et~al.}(1998){Hauser}, {Arendt}, {Kelsall}, {Dwek},
  {Odegard}, {Weiland}, {Freudenreich}, {Reach}, {Silverberg}, {Moseley},
  {Pei}, {Lubin}, {Mather}, {Shafer}, {Smoot}, {Weiss}, {Wilkinson}, \&
  {Wright}}]{Hauser1998}
{Hauser}, M.~G., {Arendt}, R.~G., {Kelsall}, T., {et~al.} 1998, \apj, 508, 25

\bibitem[{{Hauser} \& {Dwek}(2001)}]{Hauser2001}
{Hauser}, M.~G. \& {Dwek}, E. 2001, \araa, 39, 249

\bibitem[{{Juvela} {et~al.}(2009){Juvela}, {Mattila}, {Lemke}, {Klaas},
  {Leinert}, \& {Kiss}}]{Juvela2009}
{Juvela}, M., {Mattila}, K., {Lemke}, D., {et~al.} 2009, \aap, 500, 763

\bibitem[{{Kashlinsky}(2005)}]{Kashlinsky2005}
{Kashlinsky}, A. 2005, \physrep, 409, 361

\bibitem[{{Krist}(2006)}]{Krist2006}
{Krist}, J. 2006, Tiny Tim/Spitzer User's Guide

\bibitem[{{Lagache} {et~al.}(1999){Lagache}, {Abergel}, {Boulanger},
  {D{\'e}sert}, \& {Puget}}]{Lagache1999}
{Lagache}, G., {Abergel}, A., {Boulanger}, F., {D{\'e}sert}, F.~X., \& {Puget},
  J.-L. 1999, \aap, 344, 322

\bibitem[{{Lagache} {et~al.}(2003){Lagache}, {Dole}, \& {Puget}}]{Lagache2003}
{Lagache}, G., {Dole}, H., \& {Puget}, J.-L. 2003, \mnras, 338, 555

\bibitem[{{Lagache} {et~al.}(2004){Lagache}, {Dole}, {Puget},
  {P{\'e}rez-Gonz{\'a}lez}, {Le Floc'h}, {Rieke}, {Papovich}, {Egami},
  {Alonso-Herrero}, {Engelbracht}, {Gordon}, {Misselt}, \&
  {Morrison}}]{Lagache2004}
{Lagache}, G., {Dole}, H., {Puget}, J.-L., {et~al.} 2004, \apjs, 154, 112

\bibitem[{{Lagache} {et~al.}(2000){Lagache}, {Haffner}, {Reynolds}, \&
  {Tufte}}]{Lagache2000}
{Lagache}, G., {Haffner}, L.~M., {Reynolds}, R.~J., \& {Tufte}, S.~L. 2000,
  \aap, 354, 247

\bibitem[{{Lagache} {et~al.}(2005){Lagache}, {Puget}, \& {Dole}}]{Lagache2005}
{Lagache}, G., {Puget}, J.-L., \& {Dole}, H. 2005, \araa, 43, 727

\bibitem[{{Le Borgne} {et~al.}(2009){Le Borgne}, {Elbaz}, {Ocvirk}, \&
  {Pichon}}]{Le_Borgne2009}
{Le Borgne}, D., {Elbaz}, D., {Ocvirk}, P., \& {Pichon}, C. 2009, \aap, 504,
  727

\bibitem[{{LeFloc'h} {et~al.}(2009){LeFloc'h}, {Aussel}, {Ilbert}, {Riguccini},
  {Frayer}, {Salvato}, {Arnouts}, {Surace}, {Feruglio}, {Rodighiero}, {Capak},
  {Kartaltepe}, {Heinis}, {Sheth}, {Yan}, {McCracken}, {Thompson}, {Sanders},
  {Scoville}, \& {Koekemoer}}]{Le_Floch2009}
{LeFloc'h}, E., {Aussel}, H., {Ilbert}, O., {et~al.} 2009, \apj, 703, 222

\bibitem[{{Marsden} {et~al.}(2009){Marsden}, {Ade}, {Bock}, {Chapin}, {Devlin},
  {Dicker}, {Griffin}, {Gundersen}, {Halpern}, {Hargrave}, {Hughes}, {Klein},
  {Mauskopf}, {Magnelli}, {Moncelsi}, {Netterfield}, {Ngo}, {Olmi}, {Pascale},
  {Patanchon}, {Rex}, {Scott}, {Semisch}, {Thomas}, {Truch}, {Tucker},
  {Tucker}, {Viero}, \& {Wiebe}}]{Marsden2009}
{Marsden}, G., {Ade}, P.~A.~R., {Bock}, J.~J., {et~al.} 2009, ArXiv e-prints

\bibitem[{{Montier} \& {Giard}(2005)}]{Montier2005}
{Montier}, L.~A. \& {Giard}, M. 2005, \aap, 439, 35

\bibitem[{{Odegard} {et~al.}(2007){Odegard}, {Arendt}, {Dwek}, {Haffner},
  {Hauser}, \& {Reynolds}}]{Odegard2007}
{Odegard}, N., {Arendt}, R.~G., {Dwek}, E., {et~al.} 2007, \apj, 667, 11

\bibitem[{{Papovich} {et~al.}(2004){Papovich}, {Dole}, {Egami}, {Le Floc'h},
  {P{\'e}rez-Gonz{\'a}lez}, {Alonso-Herrero}, {Bai}, {Beichman}, {Blaylock},
  {Engelbracht}, {Gordon}, {Hines}, {Misselt}, {Morrison}, {Mould},
  {Muzerolle}, {Neugebauer}, {Richards}, {Rieke}, {Rieke}, {Rigby}, {Su}, \&
  {Young}}]{Papovitch2004}
{Papovich}, C., {Dole}, H., {Egami}, E., {et~al.} 2004, \apjs, 154, 70

\bibitem[{{Pascale} {et~al.}(2009){Pascale}, {Ade}, {Bock}, {Chapin}, {Devlin},
  {Dye}, {Eales}, {Griffin}, {Gundersen}, {Halpern}, {Hargrave}, {Hughes},
  {Klein}, {Marsden}, {Mauskopf}, {Moncelsi}, {Netterfield}, {Olmi},
  {Patanchon}, {Rex}, {Scott}, {Semisch}, {Thomas}, {Truch}, {Tucker},
  {Tucker}, {Viero}, \& {Wiebe}}]{Pascale2009}
{Pascale}, E., {Ade}, P.~A.~R., {Bock}, J.~J., {et~al.} 2009, ArXiv e-prints

\bibitem[{{Pearson} \& {Khan}(2009)}]{Pearson2009}
{Pearson}, C. \& {Khan}, S.~A. 2009, \mnras, 399, L11

\bibitem[{{Peebles}(1980)}]{Peebles1980}
{Peebles}, P.~J.~E. 1980, {The large-scale structure of the universe}

\bibitem[{{Puget} {et~al.}(1996){Puget}, {Abergel}, {Bernard}, {Boulanger},
  {Burton}, {Desert}, \& {Hartmann}}]{Puget1996}
{Puget}, J.-L., {Abergel}, A., {Bernard}, J.-P., {et~al.} 1996, \aap, 308, L5+

\bibitem[{{Renault} {et~al.}(2001){Renault}, {Barrau}, {Lagache}, \&
  {Puget}}]{Renault2001}
{Renault}, C., {Barrau}, A., {Lagache}, G., \& {Puget}, J.-L. 2001, \aap, 371,
  771

\bibitem[{{Rieke} {et~al.}(2004){Rieke}, {Young}, {Engelbracht}, {Kelly},
  {Low}, {Haller}, {Beeman}, {Gordon}, {Stansberry}, {Misselt}, {Cadien},
  {Morrison}, {Rivlis}, {Latter}, {Noriega-Crespo}, {Padgett}, {Stapelfeldt},
  {Hines}, {Egami}, {Muzerolle}, {Alonso-Herrero}, {Blaylock}, {Dole}, {Hinz},
  {Le Floc'h}, {Papovich}, {P{\'e}rez-Gonz{\'a}lez}, {Smith}, {Su}, {Bennett},
  {Frayer}, {Henderson}, {Lu}, {Masci}, {Pesenson}, {Rebull}, {Rho}, {Keene},
  {Stolovy}, {Wachter}, {Wheaton}, {Werner}, \& {Richards}}]{Rieke2004}
{Rieke}, G.~H., {Young}, E.~T., {Engelbracht}, C.~W., {et~al.} 2004, \apjs,
  154, 25

\bibitem[{{Rodighiero} {et~al.}(2006){Rodighiero}, {Lari}, {Pozzi},
  {Gruppioni}, {Fadda}, {Franceschini}, {Lonsdale}, {Surace}, {Shupe}, \&
  {Fang}}]{Rodighiero2006}
{Rodighiero}, G., {Lari}, C., {Pozzi}, F., {et~al.} 2006, \mnras, 371, 1891

\bibitem[{{Rowan-Robinson}(2009)}]{Rowan2009}
{Rowan-Robinson}, M. 2009, \mnras, 394, 117

\bibitem[{{Serjeant} {et~al.}(2004){Serjeant}, {Mortier}, {Ivison}, {Egami},
  {Rieke}, {Willner}, {Rigopoulou}, {Alonso-Herrero}, {Barmby}, {Bei}, {Dole},
  {Engelbracht}, {Fazio}, {Le Floc'h}, {Gordon}, {Greve}, {Hines}, {Huang},
  {Misselt}, {Miyazaki}, {Morrison}, {Papovich}, {P{\'e}rez-Gonz{\'a}lez},
  {Rieke}, {Rigby}, \& {Wilson}}]{Serjeant2004}
{Serjeant}, S., {Mortier}, A.~M.~J., {Ivison}, R.~J., {et~al.} 2004, \apjs,
  154, 118

\bibitem[{{Shupe} {et~al.}(2008){Shupe}, {Rowan-Robinson}, {Lonsdale}, {Masci},
  {Evans}, {Fang}, {Oliver}, {Vaccari}, {Rodighiero}, {Padgett}, {Surace},
  {Xu}, {Berta}, {Pozzi}, {Franceschini}, {Babbedge}, {Gonzales-Solares},
  {Siana}, {Farrah}, {Frayer}, {Smith}, {Polletta}, {Owen}, \&
  {P{\'e}rez-Fournon}}]{Shupe2008}
{Shupe}, D.~L., {Rowan-Robinson}, M., {Lonsdale}, C.~J., {et~al.} 2008, \aj,
  135, 1050

\bibitem[{{Skrutskie} {et~al.}(2006){Skrutskie}, {Cutri}, {Stiening},
  {Weinberg}, {Schneider}, {Carpenter}, {Beichman}, {Capps}, {Chester},
  {Elias}, {Huchra}, {Liebert}, {Lonsdale}, {Monet}, {Price}, {Seitzer},
  {Jarrett}, {Kirkpatrick}, {Gizis}, {Howard}, {Evans}, {Fowler}, {Fullmer},
  {Hurt}, {Light}, {Kopan}, {Marsh}, {McCallon}, {Tam}, {Van Dyk}, \&
  {Wheelock}}]{Skrutskie2006}
{Skrutskie}, M.~F., {Cutri}, R.~M., {Stiening}, R., {et~al.} 2006, \aj, 131,
  1163

\bibitem[{{Stansberry} {et~al.}(2007){Stansberry}, {Gordon}, {Bhattacharya},
  {Engelbracht}, {Rieke}, {Marleau}, {Fadda}, {Frayer}, {Noriega-Crespo},
  {Wachter}, {Young}, {M{\"u}ller}, {Kelly}, {Blaylock}, {Henderson},
  {Neugebauer}, {Beeman}, \& {Haller}}]{Stansberry2007}
{Stansberry}, J.~A., {Gordon}, K.~D., {Bhattacharya}, B., {et~al.} 2007, \pasp,
  119, 1038

\bibitem[{{Starck} {et~al.}(1999){Starck}, {Aussel}, {Elbaz}, {Fadda}, \&
  {Cesarsky}}]{Starck1999}
{Starck}, J.~L., {Aussel}, H., {Elbaz}, D., {Fadda}, D., \& {Cesarsky}, C.
  1999, \aaps, 138, 365

\bibitem[{{Stecker} \& {de Jager}(1997)}]{Stecker1997}
{Stecker}, F.~W. \& {de Jager}, O.~C. 1997, \apj, 476, 712

\bibitem[{{Stetson}(1987)}]{Stetson1987}
{Stetson}, P.~B. 1987, \pasp, 99, 191

\bibitem[{{Szapudi}(1998)}]{Szapudi1998}
{Szapudi}, I. 1998, \apj, 497, 16

\bibitem[{{Valiante} {et~al.}(2009){Valiante}, {Lutz}, {Sturm}, {Genzel}, \&
  {Chapin}}]{Valiante2009}
{Valiante}, E., {Lutz}, D., {Sturm}, E., {Genzel}, R., \& {Chapin}, E. 2009,
  ArXiv e-prints

\bibitem[{{Wall} \& {Jenkins}(2003)}]{Wall_book}
{Wall}, J.~V. \& {Jenkins}, C.~R. 2003, {Practical Statistics for Astronomers}

\bibitem[{{Wang} {et~al.}(2006){Wang}, {Cowie}, \& {Barger}}]{Wang2006}
{Wang}, W.-H., {Cowie}, L.~L., \& {Barger}, A.~J. 2006, \apj, 647, 74

\bibitem[{{Werner} {et~al.}(2004){Werner}, {Gallagher}, \&
  {Irace}}]{Werner2004}
{Werner}, M.~W., {Gallagher}, D.~B., \& {Irace}, W.~R. 2004, Advances in Space
  Research, 34, 600

\end{thebibliography}

\begin{appendix}

\section{Uncertainties on number counts including clustering}

\label{section:unc_clustering}

\subsection{Counts-in-cells moments}

We consider a clustered population with a surface density $\rho$. The expected number of objects in a field of the size $\Omega$ is $\overline{N} = \rho \Omega$. In the Poissonian case, the standard deviation around this value is $\sqrt{N}$. For a clustered distribution, the standard deviation $\sigma_N$ is given by \citep{Wall_book}

\begin{equation}
\sigma_N = \sqrt{y.\bar{N}^2 + \bar{N}}.
\label{cinceq}
\end{equation}

The expected value of $y$ is given by \citep{Peebles1980}
\begin{equation}
y = \frac{\int_{field} \int_{field} w(\theta) d\Omega_{1} d\Omega_{2}}{\Omega^2},
\end{equation}
where $w(\theta)$ is the angular two points auto correlation function of the sources.

\subsection{Measuring source clustering as a function of flux density}

We assume the classical power law description $w(\theta) = A(S,\lambda) \theta^{1-\gamma}$ with an index $\gamma = 1.8$. So, $y$ depends only on A and on the shape of the field:
\begin{equation}
y = A(S,\lambda) \frac{\int_{field} \int_{field} \theta^{1-\gamma} d\Omega_{1} d\Omega_{2}}{\Omega^2}.
\label{cinceq2}
\end{equation}
The uncertainty on $y$ is given by \citep{Szapudi1998}:
\begin{equation}
	\sigma_{y} = \sqrt{\frac{2}{N_{cell} \bar{N}^2}}.
\end{equation}

\begin{figure}
\centering
\includegraphics{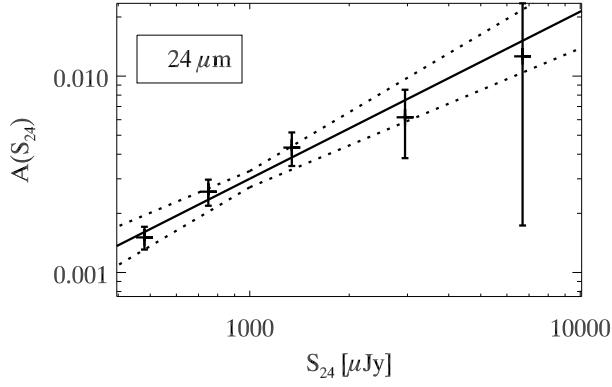}
\caption{\label{fig:fitAS} Amplitude of the auto correlation as a function of the flux density of the sources at 24 $\mu$m, and best power-law fit.}
\end{figure}

To measure $A(S,\lambda)$, we cut our fields in $30' \times 30'$ square boxes, in which we count the number of sources and compute the variance in five, three and three flux density bins at 24~$\mu$m, 70 ~$\mu$m and 160~$\mu$m. We calculate the associate $A(S,\lambda)$ combining Eq. \ref{cinceq} and \ref{cinceq2}
\begin{equation}
	A(S,\lambda) = \frac{\sigma_N^2-\bar{N}}{\bar{N}^2} \times \frac{\Omega^2}{\int_{field} \int_{field} \theta^{1-\gamma} d\Omega_{1} d\Omega_{2}}.
\end{equation}
The fit of $A(S_{24},24 \mu m)$ versus $S_{24}$ (see Fig. \ref{fig:fitAS}) gives ($\chi^2 = 2.67$ for five points and two fitted parameters)
\begin{equation}
	A(S,24 \mu m) = (2.86 \pm  0.29).10^{-3} \left (\frac{S}{1 mJy} \right )^{0.90 \pm 0.15}.
	\label{fitres}
\end{equation}
The measured exponent in \ref{fitres} of $0.90 \pm 0.15$ corresponds to $\gamma/2$, which is the expected value in the case of a flux-limited survey in an Euclidean universe filled with single luminosity sources. We fix this exponent to fit $A(S_{70}, 70 \mu m)$ and $A(S_{160}, 160 \mu m)$. We find $A(1 mJy, 70 \mu m) = (0.25 \pm 0.08 ).10^{-3}$ and $A(1 mJy, 160 \mu m) = (0.3 \pm 0.03).10^{-3}$.\\

\subsection{Compute uncertainties due to clustering}

With this model of $A(S,\lambda)$ and the field shape, we compute $y$ (Eq. \ref{cinceq2}). Assuming $\bar{N} = N$ (N to be the number of detected sources in a given field and flux density bin), we deduce $\sigma(N)$ from Eq.~\ref{cinceq}, and consequently the error bar on the number counts for a single field.\\

To compute the final uncertainty on the combined counts, we use the following relation
\begin{equation}
\sigma_{comb,\frac{dN}{dS}} = \frac{\sqrt{\sum_i \Omega_i^2 \sigma_{i,\frac{dN}{dS}}^2}}{\sum_i \Omega_i},
\end{equation}
where $\sigma_{comb,\frac{dN}{dS}}$ is the uncertainty on the combined number counts, $\Omega_i$ the solid angle of the i-th field, and $\sigma_{i,\frac{dN}{dS}}$ the uncertainty on the number counts in the i-th field (given by $Var(N)$, Eq. \ref{cinceq}).\\

\begin{figure}
\centering
\includegraphics{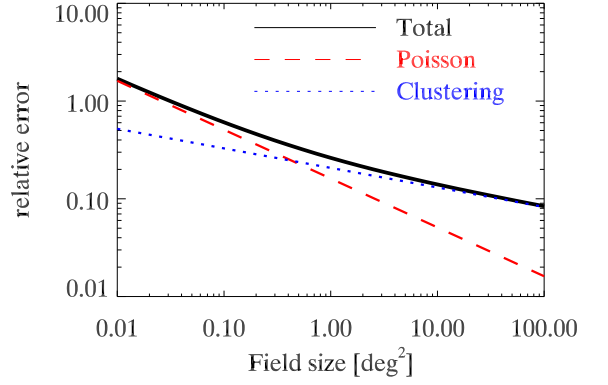}
\caption{\label{fig:relerror} Relative error on the number count as a function of the field size. We have chosen $A = 0.019$ and $\rho = 38.5$~deg$^{-2}$ (values for a 80-120~mJy flux density bin at 160 $\mu$m). The field is a square.}
\end{figure}

\subsection{Discussion about clustering and number count uncertainties}
For a clustered distribution of sources, the uncertainties on the number counts are driven by two quadratically combined terms (Eq. \ref{cinceq}): a Poissonian term $\sqrt{N}$ and a clustering term $\sqrt{y}.N$ (see Fig. \ref{fig:relerror}). We have $N \propto \Omega$ and $y \propto \Omega^{(\gamma-1)/2}$ \citep{Blake2002}. When the uncertainty is dominated by the Poissonian term (small field), the relative uncertainty is thus proportional to $\sqrt{\Omega}^{-1/2}$. When the uncertainty is dominated by the clustering term (large field), the relative error is proportional to $\Omega^{(1-\gamma)/4}$ ($\Omega^{-0.2}$ for $\gamma = 1.8$).\\

Consequently uncertainties decrease very slowly in the the clustering regime. Averaging many small independent fields gives more accurate counts than a big field covering the same surface. For example, in the clustering regime, if a field of 10~deg$^2$ has a relative uncertainty of 0.2, the relative uncertainty is 0.2/$\sqrt{10} = 0.063$ for the mean of ten fields of this size, and $0.2 \times 10^{-0.2} = 0.126$ for a single field of 100~deg$^2$. Consequently, if one studies the counts only, many small fields give better results than one very large field. But, this is not optimal if one studies the spatial properties of the galaxies, which requires large fields.\\ 

\end{appendix}

\end{document}